\documentclass[11pt]{article}
\usepackage{natbib}
\usepackage{amssymb,amsmath,amsthm,latexsym,amsfonts,amscd,dsfont,enumerate}
\usepackage{a4wide,graphicx,tikz}
\usetikzlibrary{arrows,shapes,shadows,fit,backgrounds}

\tikzstyle{trader} = [circle, draw, top color=white, bottom color=blue!30, draw=blue!50!black!100, drop shadow, minimum height=4em]
\tikzstyle{bank} = [rectangle, draw, top color=white, bottom color=red!20, draw=red!50!black!100, drop shadow, rounded corners, minimum height=3em, text width=4em, text centered]
\tikzstyle{market} = [rectangle, draw, top color=white, bottom color=green!20, draw=green!50!black!100, drop shadow, rounded corners, minimum height=3em, text width=4em, text centered]
\tikzstyle{background} = [rectangle,fill=gray!10, inner sep=0.2cm, rounded corners=5mm]
\tikzstyle{line} = [draw, latex'-latex']
\tikzstyle{from} = [draw, latex'-]
\tikzstyle{to} = [draw, -latex']

\usepackage[]{hyperref}
 \hypersetup{
 colorlinks=true, 
 breaklinks=true, 
 urlcolor= blue, 
 linkcolor= blue, 
 bookmarksopen=true, 
 pdfauthor={Pallavicini, A., Perini, D., Brigo, D.}
 }

\newcommand{\ind}[1]{1_{\{#1\}}}
\newcommand{\Exx}{\mathbb{E}}
\newcommand{\Ex}[2]{\mathbb{E}_{#1}\!\left[\,#2\,\right]}
\newcommand{\ExT}[3]{\mathbb{E}_{#1}^{#2}\!\left[\,#3\,\right]}
\newcommand{\Qxx}{\mathbb{Q}}

\newcommand{\rec}{R}
\newcommand{\lgd}{\mbox{L{\tiny GD}}}

\newtheorem{theorem}{Theorem}[section]

\newtheorem{corollary}{Corollary}[section]

\newtheorem{remark}[theorem]{Remark}

\begin{document}

\title{Funding, Collateral and Hedging:\\uncovering the mechanics and the subtleties of\\funding valuation adjustments}
\author{
Andrea Pallavicini\thanks{Financial Engineering -- Banca IMI, Milan}$\;^{,\ddag}$
\and
Daniele Perini\thanks{Financial Engineering -- Mediobanca, Milan}
\and
Damiano Brigo\thanks{Dept. of Mathematics -- Imperial College, London}
}
\date{
First version May 9, 2012.\emph{\medskip \medskip\ }This version \today.
}

\maketitle

\begin{abstract}

The main result of this paper is a collateralized counterparty valuation adjusted pricing equation, which allows to price a deal while taking into account credit and debit valuation adjustments (CVA, DVA) along with margining and funding costs, all in a consistent way. Funding risk breaks the bilateral nature of the valuation formula. We find that the equation has a recursive form, making the introduction of a purely additive funding valuation adjustment (FVA) difficult. Yet, we can cast the pricing equation into a set of iterative relationships which can be solved by means of standard least-square Monte Carlo techniques. As a consequence, we find that identifying funding costs and debit valuation adjustments is not tenable in general, contrary to what has been suggested in the literature in simple cases. The assumptions under which funding costs vanish are a very special case of the more general theory.  

We define a comprehensive framework that allows us to derive earlier results on funding or counterparty risk as a special case, although our framework is more than the sum of such special cases. We derive the general pricing equation by resorting to a risk-neutral approach where the new types of risks are included by modifying the payout cash flows. We consider realistic settings and include in our models the common market practices suggested by ISDA documentation, without assuming restrictive constraints on margining procedures and close-out netting rules. In particular, we allow for asymmetric collateral and funding rates, and exogenous liquidity policies and hedging strategies. Re-hypothecation liquidity risk and close-out amount evaluation issues are also covered.  Finally, relevant examples of non-trivial settings illustrate how to derive known facts about discounting curves from a robust general framework and without resorting to ad hoc hypotheses.

\end{abstract}

\medskip

\noindent\textbf{AMS Classification Codes}: 62H20, 91B70 \newline

\noindent \textbf{keywords}: Funding Cost, Cost of Funding, Bilateral Counterparty Risk, Credit Valuation Adjustment, Debit Valuation Adjustment, Collateral Modeling, Margining Cost, Close-Out, Re-hypothecation, Default Correlation, Gap Risk, Central Counterparty.

\pagestyle{myheadings} \markboth{}{{\footnotesize  A. Pallavicini, D. Perini and D. Brigo. Funding, Collateral and Hedging}}

\newpage
\tableofcontents
\newpage

\section{Introduction}

Cost of funding has become a paramount topic in the industry. One just has to look at the number of presentations and streams at modeling conferences of 2011-2012 that deal with this topic to realize how much research effort is being put into it. And yet literature is still in its infancy and there is very little published material, which is mostly anecdotal and not general.  

Funding costs are linked to collateral modeling, which in turn has a strong impact on credit and debit valuation adjustments (CVA and DVA). While there are several papers that try to deal with these effects separately, very few try to build a consistent framework where all such aspects can exist  together in a consistent way. Our aim in this paper is to build such a framework. 

\subsection{High Level Features}
\label{sec:intro}

We list the high level features of our theory to better clarify the motivation of our paper. Then, we move on introducing more specific aspects. 

\subsubsection{A Risk Neutral Approach}

We adopt a risk neutral approach\footnote{We are grateful to St\'ephane Cr\'epey for helpful correspondence on this issue.}. Our theory is essentially risk neutral and we do not add other bank accounts, i.e. other locally risk free assets that evolve according to different rates (see for example \cite{Crepey2011}). In a free market this would immediately lead to arbitrage, but one can live with different bank accounts if one assumes market segmentation. Rather than assuming that funding risk is embedded into different discounting and measures paradigms from the start, we add collateral, treasury and hedgers funding fees as explicit cash flows (such as for example $\varphi$). Different discounting and different measures (such as for example $\mathbb{E}^{\tilde{f}}$) may emerge later, but simply as computational tools, not as real financial issues. Our paper offers a clear explanation of the fact that we do not need to change the pricing theory, that remains the classical risk neutral theory, to account for funding costs. Our key message is that one does not need to change the theory but just the payout\footnote{This implicitly responds also to the current FVA debate on the press, see for example \cite{HW2012a,HW2012b}. In the light of our general result, already published in 2011 in \cite{Perini2011}, this looks like a storm in a teacup. We will show that a hypothetical FVA vanishes only under very special and unrealistic assumptions, including a quite inert role of the treasury in the bank, see Remark \ref{rem:dvafva}.}.

\subsubsection{Overnight Rates are not Risk-Free Rates}

Another high level feature of our approach is that the risk-free instantaneous spot rate $r_t$ associated with the bank account is not forced to be an approximate market rate.  Our distinction between market and theoretical quantities is sharp and we never claim the (false) fact that overnight rates (such as EONIA) are risk free. The risk free rate is always $r_t$ and is just an instrumental variable that vanishes if funding and hedging are accomplished through concrete market instruments, as we shall see.

\subsubsection{Keeping Treasury and Trading Activities Distinct}

A different high level aspect of our approach is that we clearly distinguish trading desks operations and treasury operations and we do not mix up such terms at closeout level (at the first default event if this happens before the final maturity of the deal) without discussing how the treasury rate emerges. The Trading and Treasury worlds are clearly connected only through the treasury rates $\tilde{f}_t$, depending in particular on the bank liquidity policy.

\subsubsection{Funding a Trade Means Funding its Hedging Strategy}

An important high level feature of our approach is that for us a price is real only if it is backed by a hedge, and the hedge is the concrete object to the trader. This is not to say that no arbitrage needs necessarily replication to be defined, since there are approaches to no-arbitrage that do not require replication. However, to back a price in practice one needs a hedge. Hence funding a product means, in our opinion, really funding its hedging strategy. This is why we put emphasis on the hedging portfolio in this paper. Of course the case where the trader has an almost perfect hedge is rare. In some cases, when the trader is not sure, she will try an approximate hedge but in general will not stand still and wait. This is why we focus on hedging. 

\subsubsection{Pricing and Perspective}

Because one entity cannot know in detail the funding policy of another entity, inclusion of funding cannot be bilateral in a valuation procedure. This means that the price of a deal between two parties will be different for the two parties. Our master formula, inclusive of funding costs, can be applied, with different funding inputs, by different parties involved in the deal, for example the bank treasury or the specific trading desk. An important point to realize is that, in general, the trader will not be able to charge its funding costs to the counterparty in the deal, and this is a further reason why the price becomes perspective-dependent. In general the trader may include funding costs into the analysis and obtain a price she may book in her system, and this price contains also the treasury costs of setting up and maintaining the deal, but the actual price that will be charged to the counterparty will not be this price. Similarly, the trader treasury may compute a different price, where the funding component is not present since this is charged to the trading desk. This is a simplistic example but helps clarifying the perspectival nature of the price we obtain. If debit valuation adjustment represents the (dubious) possibility to maintain agreement on the price of a deal when including credit risk, with funding the symmetry breaks for good and there is no longer a point in pursuing a unique price for the two parties. Different approaches can be used, but a general characterization of the relevant equilibrium is beyond the scope of this paper. 

\subsection{Single-Deal (Micro) vs. Homogeneous (Macro) Funding Models}

When dealing with funding costs, one has to first make a decision on whether to take a single deal (micro) or homogeneous (macro) cost view.

The micro approach can deal with funding costs that are deal specific. It is also an approach that distinguishes between funding and investing in terms of returns since in general different spreads will be applied when borrowing or lending. This is not the unique possibility, however, and typically treasury departments in banks work differently. One could assume an average cost of funding (borrowing) to be applied to all deals, and an average return for investing (lending). This macro approach would lead to two curves that would hold for all funding costs and invested amounts respectively, regardless of the specific deal, but depending on the sign of the exposures.

One can go further with the homogeneity assumption and assume that the cost of investing and the cost of funding match, so that spreads are not only the same across deals, but are equal to each other, implying a common funding (borrowing) and investing (lending) macro spread. In practice the spread would be set by the treasury at a common value for borrowing or lending, and this value would match what is expected to go on across all deals on average. This homogeneous average approach would look at a unique funding/lending spread for the bank treasury to be applied to all products traded by capital markets. The homogeneous approach is assumed for example in initial works such as \cite{Fries2010} and \cite{Piterbarg2010}.

In this paper we complete the work of \cite{Perini2011}, and we stay as general as possible and therefore assume a micro view, but of course it is enough to collapse our variables to common values to obtain any of the large pool approaches.    

We also point out that the micro {\it vs.} macro approaches view the treasury department in very different ways. In the micro model the treasury takes a very active role in looking at funding costs and becomes an operations center. In the homogeneous model the treasury takes more the role of a central supporting department for the bank operations. The second view is prevailing at the moment, but we point out that it is more difficult to implement absence of arbitrage in that framework.  

\subsection{Previous Literature on Funding and Collateral}

The fundamental impact of collateralization on default risk and on CVA and DVA has been analyzed in \cite{Cherubini}. More recently in \cite{BrigoCapponiPallavicini,BrigoCapponiPallaviciniPapatheodorou} CVA, DVA, and gap risk are analyzed under several collateralization strategies, with or without re-hypothecation, as a function of the margining frequency, with wrong-way risk and with possible instantaneous contagion. Minimum threshold amounts and minimum transfer amounts are also considered. We cite also \cite{Brigo2011} for a list of frequently asked questions on the subject.

The fundamental funding implications in presence of default risk have been considered in \cite{MoriniPrampolini2011}, see also \cite{Castagna2011}. These works focus on particularly simple products, such as zero coupon bonds or loans, in order to highlight some essential features of funding costs. \cite{Fujii2010} analyzes implications of currency risk for collateral modeling. 

An initial stylized analysis of the problem of replication of derivative transactions under collateralization but without default risk and in a purely classical Black and Scholes framework has been considered in \cite{Piterbarg2010}. Introduction of collateral modeling in a world without default risk is questionable, since the main role of collateral is indeed serving as a guarantee against such risk. 

The above references constitute a beginning for the funding cost literature but do not have the level of generality needed to include all the above features in a consistent framework that can then be used to manage complex products. A general theory is still missing. The only exceptions so far are \cite{BurgardKjaer2011a, BurgardKjaer2011b}, who however do not deal with the hidden complexities of collateral modeling and mark-to-market discontinuities at default. Furthermore, by resorting to a PDE approach, \cite{BurgardKjaer2011a, BurgardKjaer2011b} are unrealistically constrained to low dimensional situations. The other general result is \cite{Crepey2011} and is more promising and general, although it does not allow for credit instruments in the basic portfolio.

Our approach to introducing a general framework takes into account our past research on bilateral counterparty risk, collateral, re-hypothecation and wrong way risk across asset classes. We then add cost of funding consistently, completing the picture and building a comprehensive general framework that includes earlier results as special cases. Here, we complete the work of \cite{Perini2011}.

\subsection{Credit and Debit Valuation Adjustment (CVA and DVA)}

Prior to 2007, counterparty credit risk was accounted for through unilateral CVA (or UCVA), see for example \cite{BrigoMasetti} for the general framework under netting, and \cite{BrigoPallavicini2007}, \cite{BrigoBakkar}, \cite{BrigoChourdakis}, \cite{BrigoMoriniTarenghi} for applications of this framework to different asset classes including interest rates, equity, credit and energy commodities, all embedding wrong-way risk.

\subsubsection{Unilateral vs. Bilateral Valuation Adjustment}

The unilateral assumption, implying the omission of the DVA term, is justified, for example, when one of the two parties in the deal can be considered as being default-free, an assumption that was applied to many financial institutions prior to 2007. However, if both parties are defaultable this method is incorrect as the valuation is asymmetric between the two parties and breaks the basic accounting principle according to which an asset for one party is a liability for another. This inconsistency is remedied with the incorporation of DVA, leading to bilateral valuation adjustment BVA = CVA - DVA, where consistency imposes the presence of first to default indicators in the two CVA and DVA terms. The first-to-default clause appears in \cite{BieleckiRutkowski2002} and was made explicit in \cite{BrigoCapponi2010} in the case of an underlying CDS. It was considered in the case of interest-rate portfolios in \cite{BrigoPallaviciniPapatheodorou} and also appears in \cite{Gregory2009}. In this context, the paper \
cite{ BrigoCapponiPallaviciniPapatheodorou} extends the bilateral theory to collateralization and re-hypothecation and \cite{BrigoCapponiPallavicini} shows cases of extreme contagion where even continuous collateralization does not eliminate counterparty risk.

A simplified approach would be to consider just the difference of unilateral CVA (UCVA) and the corresponding unilateral DVA (UDVA), neglecting first to default risk. The combination BVA = UCVA - UDVA would define a simplified but inconsistent version of  bilateral valuation adjustments. The error in this simplification is analyzed in \cite{BrigoBuescuMorini}.

In 2009 ISDA modified the wording of the close-out rule in standard CSA agreements, see \cite{ISDA2009}, allowing for the possibility to switch from a risk-free close-out rule to a replacement criterion that, after the first default, includes the UDVA of the surviving party into the recoverable amount. Depending on the type of close-out that is adopted and on the default dependency between the two parties in the deal, this may lead to different types of inconsistencies highlighted for example in \cite{BrigoMorini2010Flux,BrigoMorini2011}.

\subsubsection{Counterparty Risk and Basel III}

The Basel III Accord prescribes that banks should compute unilateral UCVA by assuming independence of exposure and default. Wrong-way risk is included through one-size-fits-all multipliers. An advanced framework could allow banks to compute the effect of wrong-way risk using their own models. Examples in \cite{BrigoPallavicini2007}, \cite{BrigoBakkar}, \cite{BrigoChourdakis}, \cite{BrigoMoriniTarenghi} indicate that the actual multiplier is quite sensitive to model calibration and market conditions, and a hypothetical advanced framework might be more prudent. Interestingly, the Basel III Accord chooses to ignore the DVA in the calculation for capital adequacy requirements, although consideration of the DVA needs to be included according to accounting standards:

\begin{quote}
[...] This CVA loss is calculated without taking into account any offsetting debit valuation adjustments which have been deducted from capital under paragraph 75.  (Basel III, page 37, July 2011 release)
\end{quote}

\begin{quote}
[...] The potential for perverse incentives resulting from profit being linked to decreasing creditworthiness means capital requirements cannot recognise [DVA] (Stefan Walter, secretary-general of the Basel Committee)
\end{quote}

\begin{quote}
[...] Because nonperformance risk (the risk that the obligation will not be fulfilled) includes the reporting entities credit risk, the reporting entity should consider the effect of its credit risk (credit standing) on the fair value of the liability in all periods in which the liability is measured at fair value under other accounting pronouncements (FAS 157)
\end{quote}

This inconsistency between capital and accounting regulation is sparking much debate in the industry. 

\subsection{Inclusion of Funding Costs}

When including funding costs one has to make a number of choices. We already pointed out above that the first choice is whether to take a bottom-up view at single deal level or a large pool view. We adopt the former because it is more general, but most of the initial literature on funding takes the latter view, in line with current operational guidelines for treasury departments. Clearly the latter view is a particular case of the former, so that we are actually dealing with both views.

\subsubsection{The Recursive Nature of Pricing Equations}

When we try and include the cost of funding in the valuation of a deal we face a difficult situation. The deal future cash flows will depend on the funding choices that will be done in the future, and pricing those cash flows today involves modeling the future funding decisions. The dependence in not additively decomposable in the same way as credit valuation and debit valuation adjustments are in the case with no funding costs.

This leads to a recursive valuation equation that is quite difficult to implement, especially when dealing with products that are path dependent, since one needs at the same time backward induction and forward simulation. The recursion has been found also with different approaches, see for example \cite{Crepey2011} and \cite{BurgardKjaer2011a}.

\subsubsection{Funding Costs are not the DVA}

In this sense it is too much to expect that funding costs can be accounted for by a simple Funding Valuation Adjustment (FVA) term. Such term can be defined formally but would not add up with CVA and DVA terms in a simple way. A further consequence of the recursive nature of funding costs is that it is in general wrong to identify them with the DVA. While this happens in some very special cases, see the debate following \cite{HW2012a}, it does not hold in general.

\subsection{Structure of the Paper}

In the following sections we calculate the price of a deal inclusive of counterparty credit and debit risk (CVA and DVA), margining costs, and funding and investing costs. We start in section \ref{sec:CFBVA} by listing the main findings. Then, we describe how to obtain them. In section \ref{sec:ISDA} we extend the results of \cite{BrigoCapponiPallaviciniPapatheodorou} to include margining costs when pricing a deal with counterparty credit risk, so that we are able to fix all the terms of the pricing equation, but for funding and investing costs. In section \ref{sec:funding} we analyze the relationships between funding, hedging and collateralization to fix the remaining terms in the pricing equation. In section \ref{sec:examples} we describe how to solve the pricing equation for three realistic settings.

\section[Collateralized Credit and Funding Valuation Adjustments]{Collateralized Credit and Funding Valuation Adjustments}
\label{sec:CFBVA}

We develop a risk-neutral evaluation methodology for the Collateral-inclusive Bilateral  (credit and debit) Valuation Adjusted (CBVA) price which we extend to cover the case of Collateral and Funding inclusive Bilateral (credit and debit) Valuation Adjusted (CFBVA) price by including funding costs. Along the way, we highlight the relevant market standards and agreements which we follow to derive such formulae. We refer the reader to \cite{BrigoCapponiPallaviciniPapatheodorou} for an extensive discussion of market considerations and of collateral mechanics, which also includes an analysis of credit valuation adjustments on interest rate swaps in presence of different collateralization strategies.

In order to price a financial product (for example a derivative contract), we have to discount all the cash flows occurring after the trading position is entered. We can group them as follows:
\begin{enumerate}
\item product cash flows (e.g. coupons, dividends, etc\dots) inclusive of hedging instruments;
\item cash flows required by the collateral margining procedure;
\item cash flows required by the funding and investing procedures;
\item cash flows occurring on default events.
\end{enumerate}%

\begin{remark}{\bf (The risk-free rate as a instrumental variable)}
Notice that we discount cash flows by using the risk-free discount factor $D(t,T)$, since all costs are included as additional cash flows rather than ad hoc spreads. Here, we adopt a risk-neutral pricing framework, so that we assume the existence of a risk-free rate. Yet, as we will show in many examples, we find that we do not need to know the risk free rate value. Indeed, when the counterparties may build their hedging strategies by funding or investing through the cash accounts at their disposal (collateral assets in case of re-hypothecation, treasury's cash accounts, repo markets, etc\dots), then the risk free rate disappears from the pricing equations. 
\end{remark}

We refer to the two names involved in the financial contract and subject to default risk as investor (also called name ``I'') and counterparty (also called name ``C''). For example, ``I'' could be a Bank and ``C'' could be a corporate client. We denote by $\tau_I$,and $\tau_C$ respectively the default times of the investor and counterparty. We fix the portfolio time horizon $T \in \mathds{R}^+$, and fix the risk-neutral pricing model $(\Omega,\mathcal{G},\mathbb{Q})$, with a filtration $(\mathcal{G}_t)_{t \in [0,T]}$ such that $\tau_C$, $\tau_I$ are $\mathcal{G}$-stopping times. We denote by $\Exx_t$ the conditional expectation under $\Qxx$ given $\mathcal{G}_t$, and by $\Exx_{\tau_i}$ the conditional expectation under $\Qxx$ given the stopped filtration $\mathcal{G}_{\tau_i}$. We exclude the possibility of simultaneous defaults, and define the first default event between the two parties as the stopping time
\[
\tau := \tau_C \wedge \tau_I.
\]

The main result of the present paper is the pricing equation (CFBVA price) for a deal inclusive of collateralized credit and debit risk (CVA and DVA), margining costs, and funding and investing costs.

The CFBVA price ${\bar V}_t$ of a product (for example a derivative contract), which is derived in the following sections, is given by
\begin{eqnarray}
\label{eq:fundingpreview}
{\bar V}_t(C,F)
& = & \Ex{t}{\Pi(t,T\wedge\tau) + \gamma(t,T\wedge\tau;C) + \varphi(t,T\wedge\tau;F) } \\\nonumber
& + & \Ex{t}{\ind{t<\tau<T} D(t,\tau) \theta_\tau(C,\varepsilon) },
\end{eqnarray}%
where
\begin{itemize}
\item $\Pi(t,T)$ is the sum of all discounted payoff terms in the interval $(t,T]$, without credit or debit risk, without funding costs and without collateral cash flows;
\item $\gamma(t,T;C)$ are the collateral margining costs cash flows within the interval $(t,T]$, $C$ being the collateral account,
\item $\varphi(t,T;F)$ are the funding and investing costs cash flows within such interval, $F$ being the cash account needed for trading, and
\item $\theta_\tau(C,\varepsilon)$ is the on-default cash flow, $\varepsilon$ being the residual value of the claim being traded at default, also interpreted as the replacement cost for the deal at default (close-out amount).
\end{itemize}

The margining procedure and the liquidity policy dictate respectively the dynamics of the collateral account $C_t$ and of the funding cash account $F_t$, while the close-out amount $\varepsilon_t$ is defined by the CSA that has been agreed between the counterparties. Common strategies, as we will see later on, may link the values of such processes to the price of the product itself, transforming the previous definition in Eq (\ref{eq:fundingpreview}) into a recursive equation. This feature is hidden in simplified approaches based on adding a spread to the discount curve to accommodate collateral and funding costs. A different approach is followed by \cite{Crepey2011} and \cite{BurgardKjaer2011a}. Especially the work in \cite{Crepey2011} extends the usual risk-neutral evaluation framework to include many cash accounts accruing at different rates. Despite the different initial approach, a structure that is similar to our result above for the derivative price is obtained as a solution of a backward SDE.

In the following sections we expand all the above terms to allow the calculation of the CFBVA price.

\section{Trading under Collateralization and Close-Out Netting}
\label{sec:ISDA}

In this section we derive the pricing equation for a financial product when taking into account credit and debit effects, collateralization and netting, but no funding yet, leading to the quantity  ${\bar V}_t(C)$ with $\varphi(t,T;F)=0$.

When dealing with collateral and netting, one refers typically to ISDA (International Swaps and Derivatives Association) documentation.  Indeed, the ISDA Master Agreement lists two different tools to reduce counterparty credit risk: collateralization by a margining procedure and close-out netting rules. Both tools are ruled by the Credit Support Annex (CSA) holding between the counterparties in the deal. 

Collateralization means the right of recourse to some asset of value that can be sold or the value of which can be applied as a guarantee in the event of default on the transaction. Close-out netting rules apply when a default occurs, and force multiple obligations towards a counterparty to be consolidated into a single net obligation before recovery is applied.

Here, we briefly describe these tools to analyze the margining costs required by the collateral posting, and to integrate their price within our subsequent funding costs model, leading eventually to calculation of the Collateral-includive (credit and debit)  Bilateral Valuation Adjusted (CBVA) price ${\bar V}_t(C)$. We address the reader to \cite{BrigoCapponiPallaviciniPapatheodorou} for more details or to \cite{BrigoMoriniPallavicini2012} for a complete discussion.

\subsection{Re-Hypothecation Liquidity Risk}

In case of no default happening, at maturity the collateral provider expects to get back the remaining collateral from the collateral taker. Similarly, in case of default happening earlier (and assuming the collateral taker before default to be the surviving party), after netting the collateral account with the cash flows of the transaction, the collateral provider expects to get back the remaining collateral on the account if any. However, it is often considered to be important, commercially, for the collateral taker to have relatively unrestricted use of the collateral until it must be returned to the collateral provider. Thus, when re-hypothecation occurs, the collateral provider must consider the possibility to recover only a fraction of his collateral, see \cite{BrigoCapponiPallaviciniPapatheodorou} for the details. Re-hypothecation is widespread as a practice, and this is understandable, since it can lower the cost of collateral remuneration. For an investigation on the extent of re-hypothecation and 
of collateral velocity see \cite{Singh2010}, \cite{Singh2011}. 

Returning to our framework, if the investor is the collateral taker, we denote the recovery fraction on the collateral re-hypothecated by the defaulted investor by $\rec'_I$, while if the counterparty is the collateral taker, then we denote the recovery fraction on collateral re-hypothecated by the counterparty by $\rec'_C$. Accordingly, we define the collateral loss incurred by the counterparty upon investor default by $\lgd'_I = 1 - \rec'_I$ and the collateral loss incurred by the investor upon counterparty default by $\lgd'_C = 1 - \rec'_C$. All such quantities are defined on a unit notional.

Typically, the surviving party has precedence on other creditors to get back her posted collateral, so that $\rec_I \leq \rec_I' \leq 1$, and $\rec_C \leq \rec_C' \leq 1$. Here, $\rec_I$ ($\rec_C$) denotes the recovery fraction of the market value of the transaction that the counterparty (investor) gets when the investor (counterparty) defaults. Notice that the case when collateral cannot be re-hypothecated and has to be kept into a segregated account is obtained by setting \[\rec_I' = \rec_C' = 1.\] We do not rule out the case  $\rec' = \rec$, where collateral losses are treated as standard unsecured debit losses upon default. 

The impact of re-hypothecation on the pricing of counterparty risk is analyzed in \cite{BrigoCapponiPallaviciniPapatheodorou} for interest-rate derivatives, and in \cite{BrigoCapponiPallavicini} for credit derivatives. 

\subsection{Collateral Management under Margining Procedures}

A margining procedure consists in a pre-fixed set of dates during the life of a deal when both parties post or withdraw collateral amounts, according to their current exposure, to or from an account held by the collateral taker. A realistic margining practice should allow for collateral posting only on a fixed time-grid (${\cal T}_c := \{t_1,\ldots,t_n\}$), and for the presence of independent amounts, minimum transfer amounts, thresholds, and so on, as described in \cite{BrigoCapponiPallaviciniPapatheodorou}. Here, we present a framework which does not rely on the particular margining procedure adopted by the counterparties.

Without loss of generality, we assume that the collateral account $C_t$ is held by the investor if $C_t>0$ (the Investor is the collateral taker), and by the counterparty if $C_t<0$ (the Counterparty is the collateral taker). If at time $t$ the investor posts some collateral we write $dC_t<0$, and the other way round if the counterparty is posting.

The CSA agreement holding between the counterparties ensures that the collateral taker remunerates the account at a particular accrual rate. We introduce the collateral accrual rates\footnote{With a slight abuse of notation we use plus and minus signs to indicate which rate is used to accrue collateral according to collateral account sign, and {\emph not} to indicate that rates are positive or negative parts of some other rate.}, namely $c^+_t(T)$ when collateral assets are taken by the investor, and $c^-_t(T)$ in the other case, as adapted processes. Furthermore, we define the (collateral) zero-coupon bonds $P^{c^\pm}_t(T)$ as given by
\[
P^{c^\pm}_t(T) := \frac{1}{1+(T-t)c^\pm_t(T)} \,.
\]%
It is also useful to introduce the effective collateral accrual rate ${\tilde c}_t$ defined as
\begin{equation}
\label{eq:ctilda}
{\tilde c}_t(T) := c^-_t(T) \ind{C_t<0} + c^+_t(T) \ind{C_t>0} \,,
\end{equation}
and the corresponding zero-coupon bond
\[
P^{\tilde c}_t(T) := \frac{1}{1+(T-t){\tilde c}_t(T)} \,.
\]%

We assume that interests accrued by the collateral account are saved into the account itself, so that they can be directly included into close-out and margining procedures. Thus, any cash-flow due to collateral costs or accruing interests can be dropped from our explicit list, since it can be considered as a flow within each counterparty.

We start by listing all cash-flows originating from the investor and going to the counterparty if default events do not occur:
\begin{enumerate}
\item The Investor opens the account at the first margining date $t_1$ if $C_{t_1}<0$ (the counterparty ``C'' is the collateral taker);
\item The Investor posts to or withdraws from the account at each $t_k$, as long as $C_{t_k}<0$ (i.e. as long as the Counterparty is the collateral taker), by considering a collateral account's growth at CSA rate $c^-_{t_k}(t_{k+1})$ between posting dates;
\item The Investor closes the account at the last margining date $t_m$ if $C_{t_m}<0$.
\end{enumerate}
The counterparty considers the same cash-flows for opposite values of the collateral account at each margining date. Hence, we can sum all such contributions. If we do not take into account default events, we define the sum of (discounted) margining cash flows occurring within the time interval $A$ with $t_a:=\inf\{A\}$ as given by
\begin{eqnarray*}
\Gamma(A;{\cal T}_c,C)
& := & \,\ind{t_1\in A} C^-_{t_1} D(t_a,t_1) - \ind{t_n\in A} C^-_{t_n} D(t_a,t_n) \\
&  - & \sum_{k=1}^{n-1} \ind{t_{k+1}\in A} \left( C^-_{t_k} \frac{1}{P^{c^-}_{t_k}(t_{k+1})} - C^-_{t_{k+1}} \right) D(t_a,t_{k+1}) \\
&  + & \,\ind{t_1\in A} C^+_{t_1} D(t_a,t_1) - \ind{t_n\in A} C^+_{t_n} D(t_a,t_n) \\
&  - & \sum_{k=1}^{n-1} \ind{t_{k+1}\in A} \left( C^+_{t_k} \frac{1}{P^{c^+}_{t_k}(t_{k+1})} - C^+_{t_{k+1}} \right) D(t_a,t_{k+1}) \,,
\end{eqnarray*}%
where in our notation (which is not standard notation concerning the negative part, but useful to preserve a simpler form for equations)
\[
X^+ := \max(X,0)
\;,\quad
X^- := \min(X,0) \,.
\]%
We can re-arrange the previous equation by summing, when possible, positive and negative parts to obtain
\begin{eqnarray*}
\Gamma(A;{\cal T}_c,C)
&  = & \sum_{k=1}^{n-1} \ind{t_k\in A} \left( C_{t_k} D(t_a,t_k) - C^-_{t_k} \frac{D(t_a,t_{k+1})}{P^{c^-}_{t_k}(t_{k+1})} - C^+_{t_k} \frac{D(t_a,t_{k+1})}{P^{c^+}_{t_k}(t_{k+1})} \right) \\
&  + & \sum_{k=1}^{n-1} \left(\ind{t_k\in A}-\ind{t_{k+1}\in A}\right) \left( C^-_{t_k} \frac{D(t_a,t_{k+1})}{P^{c^-}_{t_k}(t_{k+1})} + C^+_{t_k} \frac{D(t_a,t_{k+1})}{P^{c^+}_{t_k}(t_{k+1})} \right) \,.
\end{eqnarray*}%

Now we calculate the previous expression in the following case: we consider the time interval $A(t,T;\tau)$ which goes from $t$ to $\tau \wedge T$, it contains $t$, and if $T<\tau$ it is closed on right, thereby containing $T$, otherwise it is open on the right, thereby not containing $\tau$. Such interval can be expressed in formula by
\[
A(t,T;\tau) := \{ u : t\le u\le T<\tau \} \cup \{ u : t\le u<\tau\le T \} = [t, \ \min(\tau^-,T)] 
\]%
with $t \le t_1$. The last representation is meant to be informal. We focus on the margining cash flows within $A(t,T;\tau)$, and we define $\bar\Gamma$ by
\begin{eqnarray}
\label{eq:Gamma}
{\bar\Gamma}(t,T;C)
& := & \Gamma(A(t,T;\tau);{\cal T}_c,C) \\\nonumber
&  = & \sum_{k=1}^{n-1} \ind{t_k<\tau} \left( D(t,t_k) C_{t_k} - D(t,t_{k+1}) \mu(t_k,t_{k+1}) \right) \\\nonumber
&  + & \sum_{k=1}^{n-1} \ind{t_k<\tau\le t_{k+1}} D(t,t_{k+1}) \mu(t_k,t_{k+1}) \,,
\end{eqnarray}%
where $\mu(t_k,t_{k+1})$ is the value of the collateral account accrued from date $t_k$ to date $t_{k+1}$ as required by the CSA holding between the investor and the counterparty, namely
\[
\mu(t_k,t_{k+1}) := \frac{C^-_{t_k}}{P^{c^-}_{t_k}(t_{k+1})} + \frac{C^+_{t_k}}{P^{c^+}_{t_k}(t_{k+1})} \,.
\]%

Hence we can take the risk-neutral expectation of both sides of equation~\eqref{eq:Gamma} to calculate the price of all margining cash flows, and we obtain
\[
\Ex{t}{{\bar\Gamma}(t,T;C)} = \Ex{t}{\gamma(t,T\wedge\tau;C) + \ind{\tau<T} D(t,\tau) C_{\tau^-} } \,,
\]%
where the margining costs $\gamma(t,T\wedge\tau;C)$ are defined as
\begin{equation}
\label{eq:gamma}
\gamma(t,T\wedge\tau;C) := \sum_{k=1}^{n-1} \ind{t_k<T\wedge\tau} D(t,t_k) C_{t_k} \left( 1 - \frac{P_{t_k}(t_{k+1})}{P^{\tilde c}_{t_k}(t_{k+1})} \right) \,,
\end{equation}%
and we introduce the pre-default value $C_{\tau^-}$ of the collateral account as given by
\begin{equation}
C_{\tau^-} := \sum_{k=1}^{n-1} \ind{t_k<\tau\le t_{k+1}} C_{t_k} \frac{P_\tau(t_{k+1})}{P^{\tilde c}_{t_k}(t_{k+1})} \,.
\end{equation}%

In the following, to simplify notation, we usually write $\ind{\tau<u}$ instead of $\ind{\tau\le u}$, since we are assuming that the probability that the default event happens at a particular time is zero. More specifically, we assume the distribution of the random variable $\tau$ to be continuous so that ${\mathbb Q}(\tau = u) = 0$ for all $u \ge 0$. 

Notice that we can safely assume that $t\le t_1$ in the present derivation of margining costs, since we are evaluating the price adjustment due to the whole collateralization procedure, comprising all margining dates. In the following, when calculating the price of the contract at a future time following $t_1$, we will not need to consider $t>t_1$ and repeat the current derivation. It will be simply enough to adjust the contract's price for the margining costs defined in \eqref{eq:gamma} and occurring after $t$.

\begin{remark}{\bf (Pre-default collateral account and re-hypothecation)}
The pre-default value $C_{\tau^-}$ of the collateral account is used by the CSA to calculate close-out netted exposures, and it can be different from the actual value of the collateral account at the default event, since some collateral assets (or all) might be re-hypothecated. Indeed, in section \ref{sec:closeout}, when applying close-out netting rules, first we will net the exposure against
$C_{\tau^-}$, then we will treat any remaining collateral as an unsecured claim.
\end{remark}

\subsection{Close-Out Netting Rules}
\label{sec:closeout}

The occurrence of a default event gives the interested parties the right to terminate all transactions that are included under the relevant ISDA Master Agreement. The ISDA Master Agreement sets forth the mechanism of close-out netting to be enforced. The surviving party should evaluate the terminated transactions to claim for a reimbursement after the application of netting rules consolidating the transactions, inclusive of collateral accounts.

The ISDA Master Agreement defines the term {\em close-out amount} to be the amount of losses or costs the surviving party would incur in replacing or in providing for an economic equivalent to the relevant transaction. Notice that the close-out amount is not a symmetric quantity w.r.t.\ the exchange of the role of the two parties, since it is valued by one party after the default of the other one.

We introduced the close-out amount in Section \ref{sec:CFBVA} as one of the elements needed to calculate on-default cash flows, and we named it $\varepsilon$. Since its value depends on which party survives, we specialize it in the following form:
\[
\varepsilon_\tau := \ind{\tau=\tau_C} \varepsilon_{I,\tau} + \ind{\tau=\tau_I} \varepsilon_{C,\tau} \,,
\]%
where we define the close-out amount priced at time $\tau$ by the investor on counterparty's default with $\varepsilon_{I,\tau}$,  and we define $\varepsilon_{C,\tau}$ as the close-out amount when the investor is defaulting. Notice that, while we recognize that Investor and Counterparty may measure close-out differently, from a point of view of the pure sign of the cash flows we always consider all prices from the point of view of the investor. Thus, we have four cases:
\begin{enumerate}
\item a positive $\varepsilon_{I,\tau}$ means that the Investor is a creditor of the Counterparty for the close-out measured by the Investor with Investor's methodology;
\item a positive $\varepsilon_{C,\tau}$ means that the Investor is a creditor of the Counterparty for the close-out measured by the Counterparty with Counterparty's methodology;
\item a negative $\varepsilon_{I,\tau}$ means that the Counterparty is a creditor of the Investor for the close-out measured by the Investor  with Investor's methodology;
\item a negative $\varepsilon_{C,\tau}$ means that the Counterparty is a creditor of the Investor for the close-out measured by the Counterparty  with Counterparty's methodology.
\end{enumerate}

We could include into the close-out amount also margining and funding costs as reported in \cite{isda2010}. Yet, the ISDA documentation is not very tight in defining how one has to calculate the close-out amounts, and it can clearly produce a wide range of results as described in \cite{Parker2009} and \cite{Weeber2009}. We refer to \cite{BrigoCapponiPallaviciniPapatheodorou} and to references therein for further discussion. Here, by following \cite{BrigoCapponiPallaviciniPapatheodorou}, with a view to extending it, we start by listing all the situations that may arise on counterparty default and investor default events. Our goal is to calculate the present value of all cash flows implied by the contract by taking into account 
\begin{itemize}
\item collateral margining operations, and 
\item close-out netting rules in case of default. 
\end{itemize}

We start by considering all possible situations which may arise at the default time of the counterparty, which is assumed to default before the investor. We have:

\begin{enumerate}

\item The investor measures a positive (on-default) exposure on counterparty default ($\varepsilon_{I,\tau_C}>0$), and some collateral posted by the counterparty is available ($C_{\tau_C-}>0$). Then, the investor exposure is reduced by netting, and the remaining collateral (if any) is returned to the counterparty. If the collateral is not enough, the investor suffers a loss for the remaining exposure. Thus, we have
\[
\ind{\tau=\tau_C<T} \ind{\varepsilon_{\tau} > 0} \ind{C_{\tau-}>0} (\rec_C(\varepsilon_{\tau} - C_{\tau-})^+ + (\varepsilon_{\tau}-C_{\tau-})^-) \,.
\]%

\item The investor measures a positive (on-default) exposure on counterparty default ($\varepsilon_{I,\tau_C}>0$), and some collateral posted by the investor is available ($C_{\tau_C-}<0$). Then, the investor suffers a loss for the whole exposure. All the collateral (if any) is returned to the investor if it is not re-hypothecated, otherwise only a recovery fraction of it is returned.
Thus, we have
\[
\ind{\tau=\tau_C<T} \ind{\varepsilon_{\tau} > 0} \ind{C_{\tau-}<0} (\rec_C \varepsilon_{\tau} - \rec'_C C_{\tau-}) \,.
\]%

\item The investor measures a negative (on-default) exposure on counterparty default ($\varepsilon_{I,\tau_C}<0$), and some collateral posted by the counterparty is available ($C_{\tau_C-}>0$). Then, the exposure is paid to the counterparty, and the counterparty gets back its collateral in full, yielding
\[
\ind{\tau=\tau_C<T} \ind{\varepsilon_{\tau} < 0} \ind{C_{\tau-}>0} (\varepsilon_{\tau} - C_{\tau-}) \,.
\]%

\item The investor measures a negative (on-default) exposure on counterparty default ($\varepsilon_{I,\tau_C}<0$), and some collateral posted by the investor is available ($C_{\tau_C-}<0$). Then, the exposure is reduced by netting and paid to the counterparty. The investor gets back its remaining collateral (if any) in full if it is not re-hypothecated, otherwise he only gets
the recovery fraction of the part of collateral exceeding the exposure, leading to
\[
\ind{\tau=\tau_C<T} \ind{\varepsilon_{\tau} < 0} \ind{C_{\tau-}<0} ((\varepsilon_{\tau} - C_{\tau-})^- + \rec'_C (\varepsilon_{\tau} - C_{\tau-})^+) \,.
\]%

\end{enumerate}

Similarly, if we consider all possible situations which can arise at the default time of the investor, and then aggregate all these cash flows, along with the ones due in case of no default, inclusive of the collateral account, we obtain, after straightforward manipulations,
\begin{equation}\label{eq:collcashflows}
\begin{split}
& {\bar \Pi}(t,T;C) := \\
& \quad \quad \,\,\! \Pi(t,T\wedge\tau) + {\bar\Gamma}(t,T;C) \\
& \quad + \ind{\tau=\tau_C<T} D(t,\tau) \ind{\varepsilon_{I,\tau}<0} \ind{C_{\tau^-}>0} (\varepsilon_{I,\tau} - C_{\tau^-}) \\
& \quad + \ind{\tau=\tau_C<T} D(t,\tau) \ind{\varepsilon_{I,\tau}<0} \ind{C_{\tau^-}<0} ((\varepsilon_{I,\tau} - C_{\tau^-})^- + \rec'_C (\varepsilon_{I,\tau} - C_{\tau^-})^+) \\
& \quad + \ind{\tau=\tau_C<T} D(t,\tau) \ind{\varepsilon_{I,\tau}>0} \ind{C_{\tau^-}>0} ((\varepsilon_{I,\tau} - C_{\tau^-})^- +
\rec_C(\varepsilon_{I,\tau} - C_{\tau^-})^+) \\
& \quad + \ind{\tau=\tau_C<T} D(t,\tau) \ind{\varepsilon_{I,\tau}>0} \ind{C_{\tau^-}<0} (\rec_C \varepsilon_{I,\tau} - \rec'_C C_{\tau^-}) \\
& \quad + \ind{\tau=\tau_I<T} D(t,\tau) \ind{\varepsilon_{C,\tau}>0} \ind{C_{\tau^-}<0} (\varepsilon_{C,\tau} - C_{\tau^-}) \\
& \quad + \ind{\tau=\tau_I<T} D(t,\tau) \ind{\varepsilon_{C,\tau}>0} \ind{C_{\tau^-}>0} ((\varepsilon_{C,\tau} - C_{\tau^-})^+ + \rec'_I (\varepsilon_{C,\tau} - C_{\tau^-})^-) \\
& \quad + \ind{\tau=\tau_I<T} D(t,\tau) \ind{\varepsilon_{C,\tau}<0} \ind{C_{\tau^-}<0} ((\varepsilon_{C,\tau} - C_{\tau^-})^+ +
\rec_I(\varepsilon_{C,\tau} - C_{\tau^-})^-) \\
& \quad + \ind{\tau=\tau_I<T} D(t,\tau) \ind{\varepsilon_{C,\tau}<0} \ind{C_{\tau^-}>0} (\rec_I \varepsilon_{C,\tau} - \rec'_I C_{\tau^-}).
\end{split}
\end{equation}%
This formula may be compressed further, as we are going to see shortly. 

\subsection{Collateral-inclusive Bilateral Valuation Adjusted Pricing (CBVA)}

We define the Collateral-inclusive (credit and debit) Bilateral Valuation Adjusted (CBVA) price ${\bar V}_t(C)$, without considering funding and investing costs, by taking the risk-neutral expectation of the previous Equation \eqref{eq:collcashflows}, and we obtain after a few straightforward additional manipulations
\begin{eqnarray}
\label{eq:bccva}
{\bar V}_t(C)
  & := & \Ex{t}{{\bar\Pi}(t,T;C)} \\\nonumber
  &  = & \Ex{t}{\Pi(t,T\wedge\tau) + \gamma(t,T\wedge\tau;C) + \ind{t<\tau<T} D(t,\tau) \theta_\tau(C,\varepsilon) } \,,
\end{eqnarray}%
where we define the on-default cash flow $\theta_{\tau}(C,\varepsilon)$ as given by
\begin{eqnarray}
\label{eq:theta}
\theta_{\tau}(C,\varepsilon)
 & := & \,\ind{\tau=\tau_C<\tau_I} \left( \varepsilon_{I,\tau} - \lgd_C (\varepsilon_{I,\tau}^+ - C_{\tau^-}^+)^+ - \lgd'_C (\varepsilon_{I,\tau}^- - C_{\tau^-}^-)^+ \right) \\\nonumber
 &  + & \,\ind{\tau=\tau_I<\tau_C} \left( \varepsilon_{C,\tau} - \lgd_I (\varepsilon_{C,\tau}^- - C_{\tau^-}^-)^- - \lgd'_I (\varepsilon_{C,\tau}^+ - C_{\tau^-}^+)^- \right) \,.
\end{eqnarray}%

This result implies that, to price a deal (without funding costs), we have to sum up three components:
\begin{enumerate}
\item deal cash flows (first term in the right-hand side of (\ref{eq:bccva})), 
\item margining costs (second term in the right-hand side of (\ref{eq:bccva})), and 
\item close-out amount reduced by the CVA/DVA contribution (third term in the right-hand side of (\ref{eq:bccva})).
\end{enumerate}

The above pricing formula holds also if the currency of the collateral is different from contract denomination provided that we change each foreign cash flow into domestic currency before considering it. For instance, we can consider the case of collaterals posted in a foreign currency, or even more elaborated cases where the collateral provider has the options of choosing the collateral currency during the contract life. See for more details \cite{Fujii2010}.

In the following sections we analyze two simple but relevant applications of the CBVA pricing formula: the case of perfect collateralization with collateral posted in domestic currency and the same case but with collateral posted in foreign currency. 

\subsection{Perfect Collateralization}
\label{subsec:perfect}

As an example of the CBVA master pricing formula we consider the case of ``perfect collateralization'', which we define as given by collateralization in continuous time, with continuous mark-to-market of the portfolio at default events, and with collateral account inclusive of margining costs at any time $u$, namely
\[
C_u \doteq \Ex{u}{\Pi(u,T) + \gamma(u,T;C)} \,,
\]%
with close-out amount evaluated as the collateral price, so that
\[
\varepsilon_{I,\tau} \doteq \varepsilon_{C,\tau} \doteq C_\tau \,.
\]%

Then, from the CBVA price equation we get
\begin{eqnarray*}
{\bar V}_t(C) & = & \Ex{t}{\Pi(t,T\wedge\tau) + \gamma(t,T\wedge\tau;C) + \ind{t<\tau<T} D(t,\tau) C_\tau } \\
              & = & \Ex{t}{\Pi(t,T) + \gamma(t,T;C) } \\
              & = & C_t.
\end{eqnarray*}%
We thus obtain that under perfect collateralization, namely collateralization in continuous time, with continuous mark-to-market of the portfolio in time (and at the default event in particular, {\emph i.e.} without instantaneous contagion), and with collateral account inclusive of margining costs at any time, we obtain
\begin{equation}
{\bar V}_t(C) = C_t \,.
\end{equation}

In this perfect collateralization case, we aim at characterizing the value of the deal in terms of collateral rate and instrument cash flows. In order to do this, we take a brief detour in discrete time and then take the limit.  

Consider two margining dates $t_k$ and $t_{k+1}$. By substituting the expression for margining cash flows we get (up to maturity)
\[
{\bar V}_{t_k}(C) = \frac{P^{\tilde c}_{t_k}(t_{k+1})}{P_{t_k}(t_{k+1})} \Ex{t_k}{ D(t_k,t_{k+1}){\bar V}_{t_{k+1}}(C) + \Pi(t_k,t_{k+1}) }
\;,\quad
{\bar V}_{t_n}(C) = 0 \,.
\]%
leading to, with $t_1=t$,
\[
{\bar V}_t(C) = \Ex{t}{ \sum_{k=1}^{n-1} \Pi(t_k,t_{k+1}) D(t,t_k) \prod_{i=1}^k \frac{P^{\tilde c}_{t_i}(t_{i+1})}{P_{t_i}(t_{i+1})} } \,.
\]%
Then, taking the limit of continuous collateralization, and using equation \eqref{eq:ctilda}, we obtain
\begin{equation}
\label{eq:contcoll}
{\bar V}_t(C) = \Ex{t}{ \int_t^T \Pi(u,u+du) \,\exp\left\{ -\int_t^u dv\, {\tilde c}_v \right\} } \,.
\end{equation}%

Hence, in case of perfect collateralization, we observe that valuation is obtained by discounting cash flows at the collateral rate ${\tilde c}_t$. In particular, the short rate $r_t$ has disappeared from our discounted payout.


\begin{remark}{\bf (Futures contracts)}
Futures contracts are settled daily by requiring the investor to hold a margin account which is marked-to-market according to the daily gains or losses on the contract, but the counterparties do not accrue interests on the margin account. Thus, if we do not consider other peculiarities of the contract, such as the initial and maintenance margin, we can apply equation \eqref{eq:contcoll} with ${\tilde c}_t = 0$, and we get
\[
{\bar V}_t^{\rm futures}(C) = \Ex{t}{ \int_t^T \Pi(u,u+du) } \,,
\]%
which reproduces the usual formula used to price Futures contracts, see \cite{Karatzas1998}.
\end{remark}

\subsection{Changing the Collateralization Currency}
\label{subsec:foreign}

In this section we modify the CBVA master equation to deal with collaterals in foreign currency. We name the collateral account expressed in domestic currency as before with the symbol $C_t$, while we define the collateral account expressed in foreign currency as $C^e_t$, so we get
\[
C_t := \chi_t C^e_t
\]%
with $\chi_t$ the exchange rate process which converts the foreign currency into the domestic one.

We can safely substitute the collateral account $C_t$ with such expression wherever we find it within the CBVA pricing equation, except in the expression for the collateral costs, where we have to take care of cash flow payment dates. Indeed, we have to rewrite equation \eqref{eq:Gamma} as
\begin{eqnarray}
\label{eq:Gamma_foreign}
{\bar\Gamma}(t,T;C)
& = & \sum_{k=1}^{n-1} \ind{t_k<\tau} \left( D(t,t_k) \chi_{t_k} C^e_{t_k} - D(t,t_{k+1}) \chi_{t_{k+1}} \mu^e(t_k,t_{k+1}) \right) \\\nonumber
& + & \sum_{k=1}^{n-1} \ind{t_k<\tau\le t_{k+1}} D(t,t_{k+1}) \chi_{t_{k+1}} \mu^e(t_k,t_{k+1}) \,,
\end{eqnarray}%
where we define $\mu^e(t_k,t_{k+1})$ as the value of the collateral account in foreign currency accrued from date $t_k$ to date $t_{k+1}$ as required by the CSA holding between the investor and the counterparty, namely
\[
\mu^e(t_k,t_{k+1}) := \frac{C^{e,-}_{t_k}}{P^{c^-}_{t_k}(t_{k+1})} + \frac{C^{e,+}_{t_k}}{P^{c^+}_{t_k}(t_{k+1})} \,.
\]%

Now, if we \emph{assume} that is possible to move to the foreign risk-neutral measure, we can change the fixing time of the exchange rate, namely
\[
\Ex{t_k}{\chi_{t_{k+1}} D(t_k,t_{k+1})} = \ExT{t_k}{e}{\chi_{t_k} D^e(t_k,t_{k+1})} = \chi_{t_k} P^e_{t_k}(t_{k+1}) \,,
\]%
where $D^e(t,T)$ and $P^e_t(T)$ are respectively the foreign risk-free discount factor and zero-coupon bond, and $\ExT{t}{e}{\cdot}$ is the foreign risk-neutral expectation. Thus, we can introduce the margining costs in domestic currency due to a collateralization in foreign currency as given by
\begin{equation}
\label{eq:gamma_foreign}
\gamma(t,T\wedge\tau;C) := \sum_{k=1}^{n-1} \ind{t_k<T\wedge\tau} D(t,t_k) \chi_{t_k} C^e_{t_k} \left( 1 - \frac{P^e_{t_k}(t_{k+1})}{P^{\tilde c}_{t_k}(t_{k+1})} \right) \,.
\end{equation}%
and use such expression in place of \eqref{eq:gamma} in our CBVA pricing equation.

In the case of perfect collateralization we can follow the approach as in the previous section, and after a straightforward calculation we  obtain the result of \cite{Fujii2010}

Hence, as a particular case of our framework, we obtain the price of a claim under perfect collateralization in a foreign currency, and under the assumption that it is possible to change measure between domestic and foreign risk-neutral measure. This is given by
\begin{equation}
\label{eq:contcollfor}
{\bar V}_t(C) = \Ex{t}{ \int_t^T \Pi(u,u+du) \, \exp\left\{ -\int_t^u dv\, ({\tilde c}_v - r^e_v + r_v) \right\} }
\end{equation}%
consistently with \cite{Fujii2010}.


\subsubsection{Settlement Liquidity Risk}

The pricing equation \eqref{eq:contcollfor} is valid for contracts collateralized in foreign currency, and it is based on the possibility of changing measure from the foreign risk-neutral measure to the domestic one. From a practical point of view this means that each counterparty can fund herself in foreign currency in the spot FX market without additional costs.

Yet, most domestic financial institutions cannot hold accounts directly with Central Banks of foreign states, so that they are forced to use the services of one or more custodian agents to hold their government and agency securities, see \cite{Bech2012}. Moreover, domestic financial institutions may suffer constraints on borrowing in the uncollateralized foreign inter-bank market, if foreign financial institutions are less willing to lend money, due to increased counterparty risk or liquidity needs, see \cite{Baba2009}.

Thus, an additional basis due to such settlement liquidity risk should be added to discount factors appearing in the pricing 
equation \eqref{eq:contcollfor}.

\subsubsection{Gap Risk in Single-Currency Contracts with Foreign-Currency Collaterals}

Counterparty credit risk may be mitigated by margining practice (CSA agreements), namely by using a collateral account as insurance against counterparty's default.  Yet, there are contracts that cannot be completely collateralized even with risk-free assets, since their mark-to-market value jumps at a default event.

In particular this happens for cross-currency derivatives, e.g. CCS, if we allow the FX rate to jump when one of the counterparties has defaulted, as in \cite{Schoenbucher2006}. Yet, even in case of a single-currency contract, if collateral assets are expressed in a foreign currency, we can encounter such problem, which might invalidate the possibility of a perfect collateralization.

\section{Funding Risk and Liquidity Policies}
\label{sec:funding}

We start our discussion by referring to a working paper of the Basel Committee, {\it``International Framework for Liquidity Risk Measurement, Standards and Monitoring''} of December 2009, that investigates market and funding liquidity issues. We will not proceed with its level of generality here, since we will be dealing mostly with pricing and funding as related to CVA. In pricing applications, modeling consistently funding costs with Credit and Debit Adjustments and collateral margining is a complex task, since it includes modeling the bank's liquidity policy, and to some extent the banking system as a whole.

In this respect, realistic funding liquidity modeling can be found in the literature, see for instance \cite{Drehmann2010}, or \cite{Brunnermeier2009} and references therein. Yet, here we resort to risk-neutral evaluation of funding costs, in a framework similar for example to the one by \cite{Crepey2011}. More examples can be found in \cite{BurgardKjaer2011a}, \cite{MoriniPrampolini2011}, \cite{Fries2010}, or \cite{Pallavicini2010}.

In practice, this means that while not addressing the details of funding liquidity modeling, we can simply introduce risk-neutral funding costs in terms of additional costs needed to complete each cash-flow transaction. We may add the cost of funding along with counterparty credit and debit valuation adjustments by collecting all costs coming from funding the trading position, inclusive of hedging costs, and collateral margining. We may have some asymmetries here, since the prices calculated by one party may differ from the same ones evaluated by the other party, since each price contains only funding costs undertaken by the calculating party.

\subsection{Funding, Hedging and Collateralization}

Without going into the details of funding liquidity modeling, we can introduce risk-neutral funding costs by considering the positions entered by the trader at time $t$ to obtain the amount of cash ($F_t>0$) needed to establish the hedging strategy, along with the positions used to invest cash surplus ($F_t<0$). If the deal is collateralized,  we include the margining procedure into the deal definition, so that we are able to evaluate also its funding costs. Notice that, if collateral re-hypothecation is allowed, each party can use the collateral account $C_t$ for funding, and, if the collateral account is large enough, we could drop all funding costs. See \cite{Fujii2010}.

In order to write an explicit formula for cash flows we need an expression for the cash amount $F_t$ to be funded or invested. Such problem is faced also in \cite{Piterbarg2010,Piterbarg2012}, \cite{BurgardKjaer2011a} and \cite{Crepey2011}. In this paper, to the best of our knowledge, we face this problem in the most general setting. We also point out that there are issues with the self financing condition used in \cite{Piterbarg2010} and \cite{BurgardKjaer2011a}  to derive the results, see for example \cite{BrigoRebuttal2012}; following the related correspondence, such issues have then been corrected in electronic versions of the \cite{Piterbarg2010} and \cite{BurgardKjaer2011a} papers\footnote{{see \tt{ http://www.risk.net/risk-magazine/technical-paper/1589992/funding-discounting-collateral}}\\ {\tt {-agreements-derivatives-pricing}} and \tt{ http://papers.ssrn.com/sol3/papers.cfm?abstract$\_$id=2109723}.}. 

We notice that the hedging strategy, perfectly replicating the product or derivative to be priced, is formed by a cash amount, namely our cash account $F_t$, and a portfolio $H_t$ of hedging instruments, so that, if the deal is not collateralized, or re-hypothecation is forbidden, we get
\[
F_t = {\bar V}_t(C,F) - H_t \,,
\]%

In the classical Black and Scholes theory, as illustrated for example in \cite{DuffieDAPT} when resorting to gain processes, dividend processes and price processes, the account $H_t$ would be the delta position in the underlying stock, whereas the hedging position $F_t$ would be the position in the risk free bank account.  

On the other hand, if re-hypothecation is allowed we can use collateral assets for funding, so that the amount of cash to be funded or invested is reduced and given by
\[
F_t = {\bar V}_t(C,F) - C_t - H_t \,,
\]%
where ${\bar V}_t(C,F)$ is the product risky price inclusive of funding, investing and hedging costs.

Notice that we obtain a {\em recursive} equation, since the product price at time $t$ depends on the funding strategy $F((t,T])$ after $t$, and in turn the funding strategy after $t$ will depend on the product price at subsequent future times. This will be made explicit in the following sections.

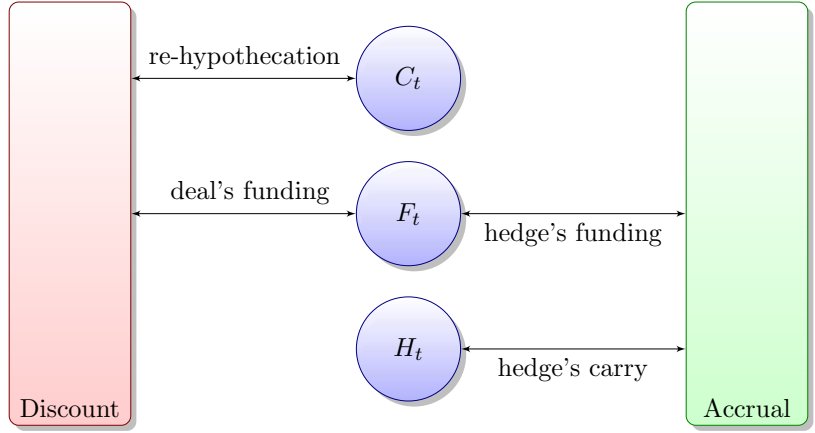
\begin{figure}
\begin{center}
\scalebox{0.9}{
\begin{tikzpicture}
   \node [trader, node distance=2cm] (C) {$C_t$};
   \node [trader, below of=C, node distance=2cm] (F) {$F_t$};
   \node [trader, below of=F, node distance=2cm] (H) {$H_t$};
   \node [bank, left of=F, text height=6cm,node distance = 5cm] (D) {Discount};
   \node [market, right of=F, text height=6cm, node distance = 5cm] (A) {Accrual};
   \path [line] (D.east)+(0,2) -- node [near end,above] {\hspace*{-1.75cm} re-hypothecation} (C);
   \path [line] (D.east)+(0,0) -- node [near end,above] {\hspace*{-1.6cm} deal's funding} (F);
   \path [line] (A.west)+(0,0) -- node [near start,below] {\hspace*{-1.8cm} hedge's funding} (F);
   \path [line] (A.west)+(0,-2) -- node [near start,below] {\hspace*{-1.8cm} hedge's carry} (H);
\end{tikzpicture}}
\end{center}
\caption{Funding, collateral, hedging and their relationships with discounting and underlying risk factors accrual rates.}
\label{fig:fch}
\end{figure}

Before going on with our analysis we try to clarify the role of the three accounts $F_t$, $C_t$, and $H_t$. When pricing a deal we have to take into account two different growth rates: discount rates and underlying risk factors' accrual rates. The former is linked to the account we obtain cash from, namely the funding cash account, and, if re-hypothecation is allowed, the collateral account. Notice that, in case re-hypothecation is forbidden, we have to separately consider the problem of funding collateral. The latter is linked on the implementation of the hedging strategy: how we fund it and the cost of carry of the hedging instruments. Indeed, the hedging instruments may require additional fees and/or collateralization. In a pictorial way we can show the relationship between the three accounts and the two accrual rates as in Figure~\ref{fig:fch}.

\subsection{Liquidity Policies}
\label{sec:liquidity}

The positions entered by the trader for funding or investing depend on his liquidity policy, namely we assume that any cash amount $F_t>0$ needed by the trader, or any cash surplus $F_t<0$ to be invested, can be managed by entering a position with an external party, for instance the treasury or a lender (``funder'') operating on the market.

In particular, we assume that the trader enters a funding position according to a time-grid $t_1,\ldots,t_m$. More precisely, between two following grid times $t_j$ and $t_{j+1}$ we have that
\begin{enumerate}
\item at $t_j$ the trader asks the funder for a cash amount equal to $F_{t_j}$;
\item at $t_{j+1}$ the trader has to reimburse the funder for the cash amount previously obtained and has to pay for funding costs.
\end{enumerate}%
Moreover, we assume that funding costs are established at the starting date of each funding period and charged at the end of the same period. We can follow the same line of reasoning also for investing cash amounts ($F_t<0$) not directly used by the trader, and consider investing periods along with funding periods.

The price of funding and investing contracts may be introduced without loss of generality as an adapted process $P^{f^+}_t(T)$, measurable at $t$, representing the price of a funding contract where the trader pays one unit of cash at maturity date $T>t$, and the price $P^{f^-}_t(T)$ of an investing contract where the trader receives one unit of cash at maturity date. We introduce also the funding and investing rates
\[
f^\pm_t(T) := \frac{1}{T-t}\left(\frac{1}{P^{f^\pm}_t(T)}-1\right).
\]%
It is also useful to introduce the effective funding and investing rate ${\tilde f}_t$ defined as
\begin{equation}
\label{eq:ftilda}
{\tilde f}_t(T) := f^-_t(T) \ind{F_t<0} + f^+_t(T) \ind{F_t>0} \,,
\end{equation}
and the corresponding zero-coupon bond
\[
P^{\tilde f}_t(T) := \frac{1}{1+(T-t){\tilde f}_t(T)} \,.
\]%

Hence, we can define the product or derivative price ${\bar V}_t(C,F)$ inclusive of funding costs and collateral management as given by
\[
{\bar V}_t(C,F) := \Ex{t}{ {\bar\Pi(t,T;C)} + \varphi(t,T\wedge\tau;F) },
\]%
where $\varphi(t,T\wedge\tau;F)$ is the sum of costs coming from all the funding and investing positions opened by the investor to hedge its trading position, according to his liquidity policy up to the first default event. This is going to be defined more precisely in a minute. Instead, ${\bar\Pi}(t,T;C)$ is the sum of the discounted cash flows coming from the product payout and inclusive of the collateral margining procedure and close-out netting rules, as given in the previous section.

Before defining $\varphi(t,T\wedge\tau;F)$, we describe a few examples of liquidity policies, in order to better illustrate our approach to funding and investing costs.

\begin{itemize}
\item[(A)] In the first case, we distinguish between funding and investing in terms of returns since there is no reason why funds lending and funds borrowing should happen at the same rate in general, so that $P^{f^-}$ and $P^{f^+}$ will be different. Moreover, the rates may also differ across deals, depending on the deals' notional, maturities structure, counterparty client relationship implications of a single product, etc. We may call this approach the ``micro'' approach to funding, or possibly the bottom-up approach. This approach is deal specific and changes rates depending on whether funds are borrowed or lent.
\item[(B)] A second possibility sees one assuming an average cost of funding borrowing to be applied to all deals, and an average return for lending or investing. This would lead to two curves for $P^{f^-}$ and $P^{f^+}$ that would hold for all funding costs and invested amounts respectively, regardless of the specific deal, so that this approach would still distinguish borrowing from lending but would not be deal specific.
\item[(C)] On the other hand, in a third approach one can go further and assume that the cost of investing and the cost of funding match, so that $P^{f^-}$ and $P^{f^+}$ are not only the same across deals, but are equal to each other, implying common funding borrowing and investing (lending) spreads. In practice the spread would be set at a common value for borrowing or lending, and this value would match what goes on across all deals on average. This would be a ``large-pool'' or homogeneous average approach, which would look at a unique funding spread for the bank to be applied to all products traded by capital markets. 
\end{itemize}

\subsubsection{Funding via the Bank's Treasury}
\label{subsec:ftp}

\begin{figure}
\begin{center}
\scalebox{0.9}{
\begin{tikzpicture}
   \node [trader, node distance=2cm] (I1) {Trader $1$};
   \node [below of=I1, node distance=2cm] (I2) {$\vdots$};
   \node [trader, below of=I2, node distance=2cm] (IN) {Trader $N$};
   \node [bank, left of=I2, text height=6cm,node distance = 5cm] (T) {Treasury};
   \node [market, left of=T, text height=6cm, node distance = 8cm] (M) {Market};
   \path [line] (T.east)+(0,2) -- node [near end,above] {$f^{+,I_1}_t$} (I1);
   \path        (T.east)+(0,2) -- node [near start,below] {$f^{-,I_1}_t$} (I1);
   \path [line] (T.east)+(0,-2) -- node [near end,above] {$f^{+,I_N}_t$} (IN);
   \path        (T.east)+(0,-2) -- node [near start,below] {$f^{-,I_N}_t$} (IN);
   \path [line] (M.east) -- node [near start,below] {$r_t+\lambda^F_t+\ell^-_t$}(T);
   \path        (M.east) -- node [near end,above] {$r_t+\lambda^I_t+\ell^+_t$}(T);
   \begin{pgfonlayer}{background}
   \node [background, fit=(I1) (IN) (T), label=above:Bank] {};
   \end{pgfonlayer}
\end{tikzpicture}}
\end{center}
\caption{Traders may borrow funds and lend/invest only by means of their treasury. Thus, average rates $f^\pm$ are applied and we are in case (B) above. Here we resort to spread notation for simplicity. $\lambda$'s are the credit spread,  and $\ell$ are the funding liquidity spreads. The funding and investing trades closed by treasury are not seen by the traders.}
\label{fig:ftp}
\end{figure}
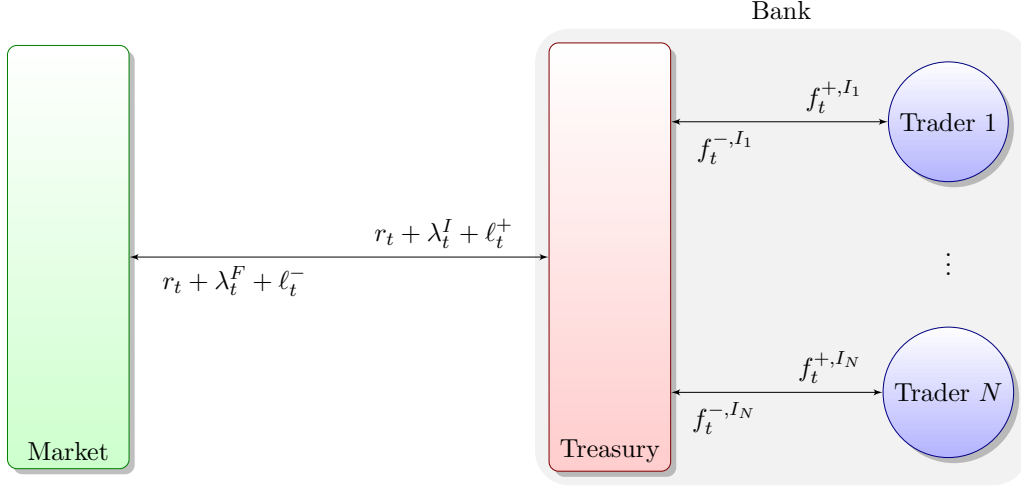

As a first example, we can consider the case where the counterparty for funding borrowing and lending (investing) cash flows is the treasury, which in turn operates on the market. Thus, funding and investing rates $f^\pm_t$ for each trader are determined by the treasury, for instance by means of a funds transfer pricing (FTP) process which allows to measure the performance of different business units. We are in case (B) above. A pictorial representation is given in figure \ref{fig:ftp}.

From the point of view of the investor the following (discounted) cash flows occur when entering a funding or investing position $\Phi_j$ at $t_j$:
\[
\Phi_j(t_j,t_{j+1};F) := - N_{t_j} D(t_j,t_{j+1}) \,,
\]%
with
\[
N_{t_j} := \frac{F^-_{t_j}}{P^{f^-}_{t_j}(t_{j+1})} + \frac{F^+_{t_j}}{P^{f^+}_{t_j}(t_{j+1})} \,.
\]%
whose price at time $t_j$ is given by $F_{t_j}$.

The Investor ``I'' does not operate directly on the market, but only with her treasury. Thus, in case of default, both the parties of the funding/investing deal disappear, without any further cash flow. In particular, in such a case the treasury, and not the trader, is in charge of debit valuation adjustments due to funding positions, so that we can consider the case where funding/investing is in place only if default events do not happen, leading to the following definition of the funding borrowing/lending (investing) cash flows when default events are considered:
\[
{\bar \Phi}_j(t_j,t_{j+1};F) := \ind{\tau>t_j} \Phi_j(t_j,t_{j+1};F) \,.
\]%
Thus, the price of cash flows coming from the $j$-th funds borrowing and lending positions is
\[
\Ex{t_j}{{\bar \Phi}_j(t_j,t_{j+1};F)} = - \ind{\tau>t_j} \left( F^-_{t_j} \frac{P_{t_j}(t_{j+1})}{P^{f^-}_{t_j}(t_{j+1})} + F^+_{t_j} \frac{P_{t_j}(t_{j+1})}{P^{f^+}_{t_j}(t_{j+1})} \right) \,.
\]%

Then we obtain that if we consider a sequence of operations to enter into a funding, borrowing and lending position at each time $t_j$ within the funding time-grid, we can define the sum $\varphi(t,T\wedge\tau;F)$ of costs coming from all the funding and investing positions opened by the investor ``I'' to hedge her trading position according to her liquidity policy up to the first default event. Such sum is given by the formula
\begin{eqnarray}
\varphi(t,T\wedge\tau;F)
&:=& \sum_{j=1}^{m-1} \ind{t\le t_j<T\wedge\tau} D(t,t_j) \left( F_{t_j} + \Ex{t_j}{{\bar \Phi}_j(t_j,t_{j+1};F)} \right) \\ \nonumber
& =& \sum_{j=1}^{m-1} \ind{t\le t_j<T\wedge\tau} D(t,t_j) \left( F_{t_j} - F^-_{t_j} \frac{P_{t_j}(t_{j+1})}{P^{f^-}_{t_j}(t_{j+1})} - F^+_{t_j} \frac{P_{t_j}(t_{j+1})}{P^{f^+}_{t_j}(t_{j+1})} \right) \,.
\end{eqnarray}
This is, strictly speaking, a payout. The cost at $t$ is obtained by taking the risk neutral expectation at time $t$ of the above cash flows. 

\subsubsection{Funding Directly on the Market}
\label{subsec:direct}

\begin{figure}
\begin{center}
\scalebox{0.9}{
\begin{tikzpicture}
   \node [trader, node distance=2cm] (I1) {Trader $1$};
   \node [below of=I1, node distance=2cm] (I2) {$\vdots$};
   \node [trader, below of=I2, node distance=2cm] (IN) {Trader $N$};
   \node [market, left of=I2, text height=6cm, node distance = 11cm] (M)
{Market};
   \path [line] (M.east)+(0,2) -- node [near end,above] {$f^{+,I_1}_t=r_t+\lambda^{I_1}_t+\ell^{+,I_1}_t$} (I1);
   \path        (M.east)+(0,2) -- node [near start,below] {$f^{-,I_1}_t=r_t+\lambda^{F_1}_t+\ell^{-,I_1}_t$} (I1);
   \path [line] (M.east)+(0,-2) -- node [near end,above] {$f^{+,I_N}_t=r_t+\lambda^{I_N}_t+\ell^{+,I_N}_t$} (IN);
   \path        (M.east)+(0,-2) -- node [near start,below] {$f^{-,I_N}_t=r_t+\lambda^{F_N}_t+\ell^{-,I_N}_t$} (IN);
   \begin{pgfonlayer}{background}
   \node [background, fit=(I1) (IN), label=above:Bank] {};
   \end{pgfonlayer}
\end{tikzpicture}}
\end{center}
\caption{Traders may borrow or lend (fund or invest) directly on the market. Thus, funding and investing rates $f^\pm$ must match the market rates. Here we resort to spread notation for simplicity. In this sense, $\lambda$'s are the default intensities of traders or funders, and $\ell^\pm$ are the liquidity (bond/CDS) bases for buying or selling. We are in case (A) above.}
\label{fig:dm}
\end{figure}
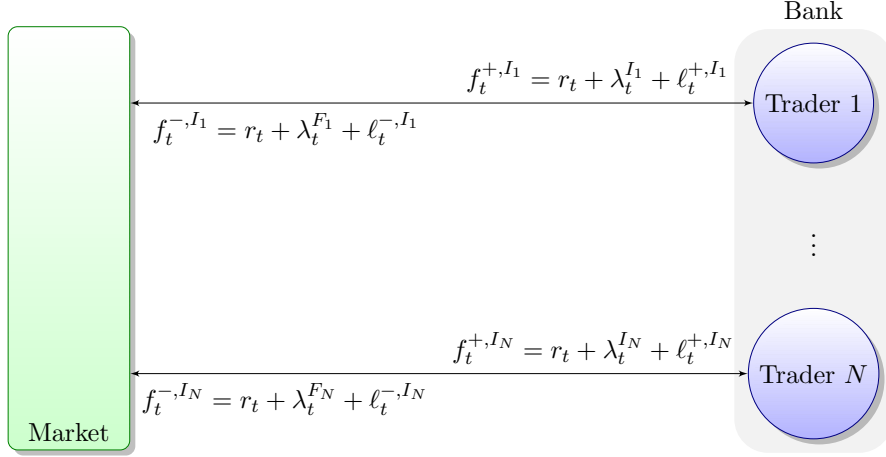

As a second example of liquidity policy, we can consider the case where each trader operates directly on the market to enter funding and investing positions (see for example \cite{Crepey2011}). Here, the treasury has no longer an active role, and it could be dropped from our scheme, as shown in Figure \ref{fig:dm}. We are in case (A) of Section \ref{sec:liquidity} above. 

The investor operates directly on the market. Thus, her mark-to-market should include the default debit valuation adjustments due to funding positions. Here, we consider the funder to be default-free.  Furthermore, we consider, as in the previous example, the case where the funding procedure is closed down if any default event happens.

By using the CBVA pricing formula without collateralization, we obtain the sum of the funding (discounted) cash flows inclusive of debit valuation adjustments as given by
\begin{eqnarray*}
{\bar\Phi}_j(t_j,t_{j+1};F)
 & := & \ind{\tau>t_j} \ind{\tau_I>t_{j+1}} \Phi_j(t_j,t_{j+1};F) \\
 &  - & \ind{\tau>t_j} \ind{\tau_I<t_{j+1}} (\lgd_I \varepsilon^-_{F,\tau_I} - \varepsilon_{F,\tau_I}) D(t_j,\tau_I) \,,
\end{eqnarray*}%
where $\varepsilon_{F,t}$ is the close-out amount calculated by the funder on investor's default event, which we assume to be
\[
\varepsilon_{F,\tau_I} := - N_{t_j} P_{\tau_I}(t_{j+1}) \,.
\]%
Notice that, by following \cite{Crepey2011}, we could assume a recovery rate for the investor different from the one we use as recovery rate for trading deals, since the seniority could be different. It is straightforward to extend the present case in such a direction.

Thus, the price of cash flows coming from the $j$-th funding and investing strategy  is given by
\[
\Ex{t_j}{{\bar \Phi}_j(t_j,t_{j+1};F)} = - \ind{\tau>t_j} \left( F^-_{t_j} \frac{P_{t_j}(t_{j+1})}{P^{f^-}_{t_j}(t_{j+1})} + F^+_{t_j} \frac{P_{t_j}(t_{j+1})}{{\bar P}^{f^+}_{t_j}(t_{j+1})} \right)\,,
\]%
where the risky-adjusted funding zero-coupon bond ${\bar P}^{f^+}_t(T)$ is defined as
\[
{\bar P}^{f^+}_t(T) := P^{f^+}_t(T) \left( \ExT{t}{T}{ \lgd_I\ind{\tau_I>T} + \rec_I } \right)^{-1}
\]%
with the expectation on the right side being taken under the $T$-forward measure.

We can now state that if we consider a sequence of operations to enter funding borrowing and lending (investing) positions at each time $t_j$ within the funding time-grid, then the sum $\varphi(t,T\wedge\tau;F)$ of costs coming from all the funding and investing positions opened by the investor to hedge its trading position according to his liquidity policy up to the first default event is given by 
\begin{eqnarray}
\varphi(t,T\wedge\tau;F)
&:=& \sum_{j=1}^{m-1} \ind{t\le t_j<T\wedge\tau} D(t,t_j) \left( F_{t_j} + \Ex{t_j}{{\bar \Phi}_j(t_j,t_{j+1};F)} \right) \\ \nonumber
& =& \sum_{j=1}^{m-1} \ind{t\le t_j<T\wedge\tau} D(t,t_j) \left( F_{t_j} - F^-_{t_j} \frac{P_{t_j}(t_{j+1})}{P^{f^-}_{t_j}(t_{j+1})} - F^+_{t_j} \frac{P_{t_j}(t_{j+1})}{{\bar P}^{f^+}_{t_j}(t_{j+1})} \right) \,.
\end{eqnarray}%
where the dependency of $\varphi$ on $\tau_I$ is not explicitly shown to avoid cumbersome notation. 

Furthermore, notice that, if we set $\rec_I=1$ (so that $\lgd_I=0$) we get that ${\bar P}^{f^+}_t(T)$ is equal to $P^{f^+}_t(T)$, and we recover the previous example. Thus, in the following we will simply write $P^{f^\pm}_t(T)$ in any case.

\subsection{CBVA Pricing with Funding Costs (CFBVA)}

In this paper we stay as general as possible and therefore assume the micro view, namely case (A) in Section \ref{sec:liquidity} above.  Of course it is enough to collapse our funding variables to common values to obtain any of the large-pool approaches (B) or (C). By the previous examples we understand that a sensible choice for funding and investing costs could be given by
\begin{equation}
\label{eq:fundinggeneral}
\varphi(t,T\wedge\tau;F) := \sum_{j=1}^{m-1} \ind{t\le t_j<T\wedge\tau} D(t,t_j) F_{t_j} \left( 1 - \frac{P_{t_j}(t_{j+1})}{P^{\tilde f}_{t_j}(t_{j+1})} \right) \,,
\end{equation}%
whatever the definition of funding and investing rates may be.

This leads to the following
\begin{theorem}{\bf (Consistent Arbitrage-Free Pricing in presence of Credit, Debit, Funding and Margining costs and adjustments).}
\label{th:thCFBVA}
The arbitrage-free Collateral and Funding inclusive (Credit-Debit-) Bilateral Valuation  Adjusted (CFBVA) price ${\bar V}_t(C,F)$, inclusive of funding and investing costs, can be written in the following form:
\begin{eqnarray}
\label{eq:CFBVA}
{\bar V}_t(C,F) & = & \Ex{t}{\Pi(t,T\wedge\tau) + \gamma(t,T\wedge\tau;C) + \varphi(t,T\wedge\tau;F) } \\\nonumber
                 & + & \Ex{t}{\ind{t<\tau<T} D(t,\tau) \theta_\tau(C,\varepsilon)
}.
\end{eqnarray}%
where $\varphi(t,T\wedge\tau;F)$ has been defined in equation \eqref{eq:fundinggeneral} and the remaining terms have been defined in Section \ref{sec:ISDA}.
\end{theorem}

The above formula, when funding and margining costs are discarded, collapses to the formula of CBVA adjusted price found in \cite{BrigoCapponiPallaviciniPapatheodorou}.

Notice that the B in CFBVA now highlights the bilateral fact that both default times of inverstor and counterparty are accounted for, but does not imply any longer that the price will be symmetric for the two parties. See the conclusions below.  

\subsection{Iterative Solution of the CFBVA Pricing Equation}
\label{sec:iterative}

In the previous sections, we derived the collateral- and funding-inclusive (credit and debit) bilateral valuation adjusted price equation \eqref{eq:CFBVA} which allows to price a deal by taking into account counterparty risk, margining and funding costs. We also built some relevant examples to highlight the recursive nature of the equation and its link with discount curves. 

Now, we describe a strategy to solve the equation without resorting to simplifying hypotheses. We try to turn the recursion into an iterative set of equations which eventually are to be solved via least-square Monte Carlo techniques (sometimes referred to as ``American Monte Carlo'') as in standard CVA calculations, see for instance \cite{BrigoPallavicini2007}.

We start by introducing the following quantities as building blocks for our iterative solution
\begin{equation}
\label{eq:barpi}
{\bar\Pi}_T(t_j,t_{j+1};C) := \Pi(t_j,t_{j+1}\wedge\tau) + \gamma(t_j,t_{j+1}\wedge\tau;C) + \ind{t_j<\tau<t_{j+1}} D(t_j,\tau) \theta_\tau(C,\varepsilon)
\end{equation}
where $\theta$ is still defined as in equation (\ref{eq:theta}). The time parameter $T$ in ${\bar\Pi}_T$ points out that the exposure $\varepsilon$ inside $\theta$ still refers to a deal with maturity $T$. From the above definition it is clear that ${\bar\Pi}_T(t,T;C) = {\bar\Pi}(t,T;C)$. 

We solve equation \eqref{eq:CFBVA} at each funding date $t_j$ in terms of the price $\bar V$ calculated at the following funding time $t_{j+1}$, and we obtain
\begin{eqnarray*}
{\bar V}_{t_j}(C;F) &=& \Ex{t_j}{ {\bar V}_{t_{j+1}}(C;F) D(t_j,t_{j+1}) + {\bar\Pi}_T(t_j,t_{j+1};C) } \\
                   &-& \ind{\tau>t_j}
                         \left( F^-_{t_j} \frac{P_{t_j}(t_{j+1})}{P^{f^-}_{t_j}(t_{j+1})}
                             + F^+_{t_j} \frac{P_{t_j}(t_{j+1})}{P^{f^+}_{t_j}(t_{j+1})}
                             - F_{t_j} \right)
\end{eqnarray*}

We recall that $F_{t_j} = {\bar V}_{t_j}(C;F) - H_{t_j}$ if re-hypothecation is forbidden, or $F_{t_j} = {\bar V}_{t_j}(C;F) - C_{t_j} - H_{t_j}$ if it is allowed. Furthermore, we have
\[
{\bar V}_{t_n}(C;F) := 0 \,.
\]

Hence, by solving for positive and negative parts, we obtain the following

\begin{theorem}
{\label{th:iter}
\bf( Iterative equations for the full pricing formula given in Theorem \ref{th:thCFBVA}).} We may express the complete pricing Formula in Theorem \ref{th:thCFBVA}), inclusive of Credit, Debit, Funding and Collateralization adjustments, in an iterative way, namely for $\tau<t_j$ we have that ${\bar V}_{t_j}(C;F)=0$, while for $\tau>t_j$:

\begin{enumerate}

\item[(i)] if re-hypothecation is forbidden, we have
\[
\left( {\bar V}_{t_j}(C;F) - H_{t_j} \right)^\pm = P^{f^\pm}_{t_j}(t_{j+1}) \left( \ExT{t_j}{t_{j+1}}{ {\bar V}_{t_{j+1}}(C;F) + \frac{{\bar\Pi}_T(t_j,t_{j+1};C)-H_{t_j}}{D(t_j,t_{j+1})} } \right)^\pm
\]

\item[(ii)] if re-hypothecation is allowed, we have
\begin{multline*}
\left( {\bar V}_{t_j}(C;F) - C_{t_j} - H_{t_j}\right)^\pm = \\ P^{f^\pm}_{t_j}(t_{j+1}) \left( \ExT{t_j}{t_{j+1}}{ {\bar V}_{t_{j+1}}(C;F) + \frac{{\bar\Pi}_T(t_j,t_{j+1};C)-C_{t_j}-H_{t_j}}{D(t_j,t_{j+1})} } \right)^\pm 
\end{multline*}

\end{enumerate}

\end{theorem}

\subsection{Dealing with Hedging Strategies}

When the investor implements a specific hedging strategy $H_t$, he may experience additional costs if he is not directly accessing the (spot) market to hedge his position in the underlying risk factors, but he is forced (or he chooses) to lend and borrow via an intermediate entity, as in the case of stock-lending and repo markets, or to trade other derivative contracts, for instance trading forward contracts on the underlying risk factors.

Indeed, prices of contracts quoted on the market may be different from prices calculated by the investor when funding and investing costs are taken into account, since investor's costs may be different from costs experienced by other market participants.\footnote{Arbitrages may appear only if market imperfections do not prevent market participants to exploit them. We consider differences in funding and investing costs as a component of usual bid-ask spreads observed on the market.} Thus, when evaluating derivative contracts, the investor has to take into account such differences in prices by including them as additional costs.

As usual we can change the drift of the price processes of underlying risk factors to include such costs. For instance, in the case of stock-lending and repo stock markets, we can use as growth rate for stock prices the quoted repo rate. Yet, if we assume that hedging costs are handled by a change of measure, we have also to drop the term depending on $H_t$ in our pricing equation.

We present in the next section a sketchy proof of the above statement in the simpler case of funding and hedging in continuous time. We leave as further work a more rigorous derivation in a more general setting.

\subsubsection{Funding Derivative Contracts in a Diffusion Setting}
\label{sub:diffusion}

Here, we solve recursively the pricing equations of Theorem~\ref{th:iter}, and take the limit of margining, funding and hedging in continuous time. We start with the case of re-hypothecation. The same approach applies also when re-hypothecation is forbidden.

When we consider all time-grids in continuous time, we get
\begin{eqnarray*}
{\bar V}_t(C;F) & = & \int_t^T \Ex{t}{ \left( \ind{u<\tau} \partial_u\pi_u + \ind{\tau\in du} \theta_u(C,\varepsilon) \right) D(t,u) } \\
                & + & \int_t^T du \, \Ex{t}{ \ind{u<\tau} ( {\tilde f}_u - {\tilde c}_u ) C_u D(t,u) } \\
                & + & \int_t^T du \, \Ex{t}{ \ind{u<\tau} ( r_u - {\tilde f}_u ) \left( {\bar V}_u(C,F) - H_u \right)D(t,u) }
\end{eqnarray*}
where $d\pi_t := \Pi(t,t+dt)$ is the coupon process.

We now consider the market filtration ${\cal F}$ that one obtains implicitly by assuming a separable structure for the filtration ${\cal
G}$, where ${\cal G}$ is generated by the pure default-free market filtration ${\cal F}$ and by the filtration generated by all the relevant default times (see for example \cite{BieleckiRutkowski2002}). Then, we switch to the market filtration ${\cal F}$, leaving the time
variable in the expectation (ie we set $\Ex{t}{X|{\cal F}}\doteq\Ex{}{X|{\cal F}_t}$):
\begin{eqnarray*}
{\bar V}_t(C;F)
& = & \ind{\tau>t} \int_t^T du \, \Ex{t}{ \left( \partial_u\pi_u + \lambda_u \theta_u(C,\varepsilon) \right) D(t,u;r+\lambda) \big| {\cal F}} \\
& + & \ind{\tau>t} \int_t^T du \, \Ex{t}{ ( {\tilde f}_u - {\tilde c}_u ) C_u D(t,u;r+\lambda) \big| {\cal F} } \\
& + & \ind{\tau>t} \int_t^T du \, \Ex{t}{ ( r_u - {\tilde f}_u ) \left( {\bar V}_u(C,F) -H_u \right) D(t,u;r+\lambda) \big| {\cal F} }
\end{eqnarray*}%
where $\lambda_t$ is the first-default intensity, and
\[
D(t,T;x) := \exp\left\{-\int_t^T du\, x_u\right\}
\]

We can write the corresponding pre-default PDE if we assume that the hypotheses of the Feynman-Kac theorem are holding, in particular that the underlying market risk factors are Markov with infinitesimal generator ${\cal L}_t$. In such case we get for $\tau>t$
\[
\left( \partial_t - {\tilde f}_t - \lambda_t + {\cal L}_t \right) {\bar V}_t(C;F) - ( r_t - {\tilde f}_t ) H_t + ( {\tilde f}_t - {\tilde c}_t ) C_t + \partial_t \pi_t = 0
\]%
with boundary condition
\[
{\bar V}_\tau(C;F) = \ind{\tau<T} \theta_\tau(C,\varepsilon)
\]

If we consider a diffusive dynamics, and we assume delta-hedging, we can expand the generator $\cal L$ in terms of first and second order operators, and we get
\[
{\cal L}_t {\bar V}_t(C;F) \doteq \left( {\cal L}_t^1 + {\cal L}_t^2 \right) {\bar V}_t(C;F) \doteq r_t H_t + {\cal L}_t^2 {\bar V}_t(C;F)
\]
Hence, the pre-default PDE becomes
\[
\left( \partial_t - {\tilde f}_t - \lambda_t + {\cal L}_t^{\tilde f} \right) {\bar V}_t(C;F) + ( {\tilde f}_t - {\tilde c}_t ) C_t + \partial_t \pi_t = 0
\]
where
\[
{\cal L}_t^{\tilde f} {\bar V}_t(C;F) := {\tilde f}_t H_t + {\cal L}_t^2 {\bar V}_t(C;F)
\]
Notice that the above equation does not depend any more on the risk-free rate. The pre-default PDE equation can be solved numerically  as in \cite{Crepey2012a}. On the other hand, we can apply again the Feynman-Kac theorem and we get the following theorem

\begin{theorem}
{\label{th:fund_hedge}
\bf( CFBVA pricing equations in case of funding and delta-hedging in continuous time).} In case of funding and delta-hedging in continuous time we may solve the iterative equations of Theorem~\ref{th:iter}, and we obtain in case of re-hypothecation
\begin{eqnarray*}
{\bar V}_t(C;F) & = & C_t + \ExT{t}{\tilde f}{ \ind{\tau<T} \left( \theta_\tau(C,\varepsilon) - C_{\tau^-} \right) D(t,\tau;{\tilde f}) } \\
                & + & \int_t^T \ExT{t}{\tilde f}{ \ind{u<\tau} \left( d\pi_u + dC_u - {\tilde c}_u C_u \,du \right) D(t,u;{\tilde f}) }
\end{eqnarray*}%

\noindent where the expectations are taken under a pricing measure $\mathbb{Q}^{\tilde f}$ under which the underlying risk factors grow at rate ${\tilde f-q}$, $q$ being the dividend yield.

\end{theorem}

We split the above pricing equation into three terms to highlight the role of different cash flows. The first term is the collateral price. The second and the third terms are formed by the cash flows which are not considered by collateralization, such as the cash flows payed on default event of one of the two counterparties, by mark-to-market movements which are not followed by an immediate reset of the collateral account, and by margining costs not included into collateral price.

A similar result holds when re-hypothecation is not allowed.

\begin{remark}\label{rem:nordep}{\bf (No explicit dependence on $r$).} An important point about our pricing equation in the above theorem is that it does not depend on the risk free rate $r_t$, which does not need to enter the modeling framework. The equation is entirely governed by market rates.
\end{remark}

\begin{remark}\label{rem:fpfmdec}{\bf (Not a real additive decomposition unless $f^+=f^-$ and further conditions hold).} Another  important point about our pricing equation is that it may appear to have achieved an additive decomposition in different adjustments if one remembers the CVA-DVA terms implicit in the closeout cash flows, including $\theta$. However, it is very important to keep in mind that the treasury rates $\tilde{f}$ future paths depend on the future signs of the account $F$, which in turn is equal to $F= \bar{V}-C-H$. This implies that future paths of treasury rates $\tilde{f}$ depend on future paths of the adjusted price $\bar{V}$ we are trying to calculate. This keeps the recursion alive and shows there is no real decomposition. One condition to move towards an additive formula is for example $f^+=f^-$, since in such case $\tilde{f}$ would no longer need to know the sign of $F$ (and hence the value of $\bar{V}$). This corresponds to the unrealistic setting where borrowing and lending can occur at the same rates. However, further issues on how the collateral is linked to $\bar{V}$ and on how the collateral rate $\tilde{c}$ may need to know the sign of $C$ show that one really needs to make sure that no quantity on the right hand side of the equation needs to know the future paths of $\bar{V}$.
\end{remark}

\subsubsection{Implementing Hedging Strategies via Derivative Markets}
\label{sub:repo}

In the previous section we considered the possibility to implement a delta-hedging strategy by trading directly on the spot market. If the investor chooses, or is forced, to trade the underlying asset by entering into derivative positions, we should add any additional cost to the CFBVA pricing equation. For instance, this situation may materialize when the investor accesses the lending/repo market to implement his hedging strategy, or if he uses synthetic forward contracts built on the European call/put market.

In general, we introduce the adapted processes $h^+_t(T)$, as the effective rates for asset lending from $t$ to $T$, and $h^-_t(T)$, for asset borrowing. Furthermore, we define the (hedging) zero-coupon bonds $P^{h^\pm}_t(T)$ as given by
\[
P^{h^\pm}_t(T) := \frac{1}{1+(T-t)h^\pm_t(T)} \,.
\]%
It is also useful to introduce the effective lending/borrowing rate ${\tilde h}_t$ defined as
\[
{\tilde h}_t(T) := h^-_t(T) \ind{H_t<0} + h^+_t(T) \ind{H_t>0} \,,
\]%
and the corresponding zero-coupon bond
\[
P^{\tilde h}_t(T) := \frac{1}{1+(T-t){\tilde h}_t(T)} \,.
\]%

Hence, if we assume that the hedging strategy is implemented on the same time-grid as the funding procedure, we can sum the funding and hedging costs in a unique term, and we can re-define $\varphi$ by explicitly taking into account its dependency on the hedging strategy.
\begin{eqnarray*}
\varphi(t,T\wedge\tau;F,H)
& \doteq & \sum_{j=1}^{m-1} \ind{t\le t_j<T\wedge\tau} D(t,t_j) F_{t_j} \left( 1 - \frac{P_{t_j}(t_{j+1})}{P^{\tilde f}_{t_j}(t_{j+1})} \right) \\
&  - & \sum_{j=1}^{m-1} \ind{t\le t_j<T\wedge\tau} D(t,t_j) H_{t_j} \left( \frac{P_{t_j}(t_{j+1})}{P^{\tilde f}_{t_j}(t_{j+1})} - \frac{P_{t_j}(t_{j+1})}{P^{\tilde h}_{t_j}(t_{j+1})} \right) \,.
\end{eqnarray*}%

If we repeat the calculation of previous sections, by taking the limit of funding and delta-hedging in continuous time, we obtain the following corollary to Theorem~\ref{th:fund_hedge}.

\begin{corollary}
{\label{th:fund_hedge_with_costs}
\bf( CFBVA pricing equations in case of funding and delta-hedging via derivative markets in continuous time).} In case of funding and delta-hedging in continuous time, when the hedging strategy is implemented by trading on a derivative market where the lending/borrowing rate is $\tilde h$, we may solve the iterative equations of Theorem~\ref{th:iter}, and we obtain in case of re-hypothecation
\begin{eqnarray*}
{\bar V}_t(C;F) & = & C_t + \ExT{t}{\tilde h}{ \ind{\tau<T} \left( \theta_\tau(C,\varepsilon) - C_{\tau^-} \right) D(t,\tau;{\tilde f}) } \\
                & + & \int_t^T \ExT{t}{\tilde h}{ \ind{u<\tau} \left( d\pi_u + dC_u - {\tilde c}_u C_u \,du \right) D(t,u;{\tilde f}) }
\end{eqnarray*}%
\noindent where the expectations are taken under a pricing measure $\mathbb{Q}^{\tilde h}$ under which the underlying risk factors grow at rate ${\tilde h}$.
\end{corollary}

\begin{remark}{\bf (Hedging in collateralized markets).}
An interesting case, where Corollary \ref{th:fund_hedge_with_costs} can be applied, is given by markets quoting only collateralized deals. A typical example is given by the money market, where liquid contracts are collateralized on a daily basis at the over-night rate. In such case, the effective lending/borrowing rate ${\tilde h}_t$ is simply given by the collateral rate itself. We refer to \cite{Pallavicini2012}, where we show how to build a multi-curve dynamical model for interest-rate derivatives by describing the market instruments in terms of martingales under a collateralized measure, for a detailed analysis.
\end{remark}

\section{Detailed Examples}
\label{sec:examples}

Here, we present three relevant examples to highlight the properties of the CFBVA pricing equation. We analyze 
\begin{itemize}
\item[(i)] the case of a perfectly collateralized contract between two risky counterparties in presence of funding and hedging costs; 
\item[(ii)] the case of a central counterparty (CCP) pricing a collateralized contract between two risky counterparties, possibly in presence of gap risk; and 
\item[(iii)] the case of a risky investor evaluating counterparty credit risk on a uncollateralized deal.
\end{itemize}

In all the following examples we consider funding and delta-hedging in continuous time, so that, according to Theorem~\ref{th:fund_hedge}, we can drop any explicit dependency on the hedging strategy, if we take all the expectations under a pricing measure $\mathbb{Q}^{\tilde f}$ under which the underlying risk factors grow at the funding rate ${\tilde f}_t$. Notice that the growth rate may be corrected for hedging costs when delta-hedging is accomplished by trading on a derivative market as stated in Corollary~\ref{th:fund_hedge_with_costs}.

\subsection{Funding with Collateral}

If re-hypothecation is allowed, we assume that we can fund with collateral assets, so that the cash amount $F_t = {\bar V}_t(C,F) - C_t - H_t$. This choice for $F_t$ leads to a recursive equation which can be solved backwards by starting from the final maturity. Notice that the collateral account value $C_t$ is defined only at margining dates, but we are taking the limiting case of perfect collateralization, so that every time is a margining date (we recall also that our definition of perfect collateralization requires that the mark-to-market of the portfolio is continuous and there is no instantaneous contagion at default in particular). Moreover, being in the re-hypothecation case, we consider recoveries as given by $\rec_C'=\rec_C$ and $\rec_I'=\rec_I$.

We assume, as in the perfect collateralization case, that we have collateralization in continuous time, with continuous mark-to-market of the portfolio in time, and with collateral account inclusive of margining costs. We recall that expectations are taken under the pricing measure $\mathbb{Q}^{\tilde f}$. Thus, we must price also the collateral account under the same measure, otherwise hedging costs must be added explicitly. Such conditions can be fulfilled by defining the collateral price as given by
\begin{equation}
\label{eq:collcloseout}
C_t \doteq \ExT{t}{\tilde f}{ \Pi(t,T) + \gamma(t,T;C) } \,.
\end{equation}%

Furthermore, the close-out amount is set equal to the collateral price, so that
\[
\varepsilon_{I,\tau} \doteq \varepsilon_{C,\tau} \doteq C_\tau \,.
\]%

Then, from Theorem~\ref{th:fund_hedge} we obtain that in the case where we may fund with collateral, under the assumptions of the present Section, we obtain a pricing formula given by
\begin{equation}
\label{eq:contfundandcoll}
{\bar V}_t(C,F) = C_t = \ExT{t}{\tilde f}{ \int_t^T \Pi(u,u+du) \, \exp\left\{ -\int_t^u dv\, {\tilde c}_v \right\} } \,,
\end{equation}%
where the expectations are taken under the $\mathbb{Q}^{\tilde f}$ pricing measure. Hence, in case of perfect collateralization, there are no funding costs, since we are funding with collateral and this has no extra cost, as shown in equation \eqref{eq:contfundandcoll}.

\subsection{Collateralized Contracts Priced by a CCP}
\label{sec:ccp}

Here, we apply our CFBVA master formula to unfunded instruments, such as interest-rate swaps or credit default swaps, which can be funded with collateral (we assume re-hypothecation is holding).

We interpret CFBVA adjusted prices as the prices calculated by a risk-free central counterparty (CCP), which interposes herself between the two counterparties of the deal. Cash flows coming from the investor to the counterparty are first payed to the CCP, which, in turn, gives them to the counterparty, and similarly for cash flows coming from the counterparty to the investor. Further, we assume that the CCP can fund herself on the money market at the overnight rate $e_t$ plus a liquidity spread $\ell^\pm_t$ (cash flows positive for the investor are negative to her).
\[
f^\pm \doteq e_t + \ell^\mp_t
\]%

In the present example we recall that expectations are taken under the pricing measure $\mathbb{Q}^{\tilde f}$, and we assume that, on a continuous time-grid, collateral is posted or withdrawn, according to the risk-free price augmented with margining given by (see (\ref{eq:collcloseout}))
\[
C_t \doteq \ExT{t}{\tilde f}{\Pi(t,T) + \gamma(t,T;C)} \,.
\]%
This leads, in the limit of continuous collateralization, via equation \eqref{eq:contfundandcoll}, to
\[
C_t = \ExT{t}{\tilde f}{ \int_t^T \Pi(u,u+du) \, \exp\left\{ -\int_t^u dv\, {\tilde c}_v \right\} } \, 
\]%
where expectations are taken under $\mathbb{Q}^{\tilde f}$ pricing measure.

Furthermore, we assume that upon a default event, the close-out amount is calculated by both counterparties in the same way as they perform the calculations for collateral assets.
\[
\varepsilon_{I,\tau} \doteq \varepsilon_{C,\tau} \doteq C_\tau
\;,\quad
C_\tau := \ExT{\tau}{\tilde f}{ \Pi(\tau,T) + \gamma(\tau,T;C)}
\]%

Moreover, we assume that funding and investing operations are entered into on a continuous time-grid

Hence, from Theorem~\ref{th:fund_hedge}, we obtain that under the assumptions of an intermediating CCP as in this Section, the adjusted product's price is given by
\begin{eqnarray*}
{\bar V}_t(C;F)
& = & \int_t^T \ExT{t}{\tilde f}{ \Pi(u,u+du) \, \exp\left\{ -\int_t^u dv\, {\tilde c}_v \right\} } \\
& - & \ExT{t}{\tilde f}{ \ind{\tau=\tau_C<T} \lgd_C (C_\tau - C_{\tau^-})^+ \, \exp\left\{ -\int_t^\tau dv\, (\ell^+_v + e_v) \right\} } \\
& - & \ExT{t}{\tilde f}{ \ind{\tau=\tau_I<T} \lgd_I (C_\tau - C_{\tau^-})^- \, \exp\left\{ -\int_t^\tau dv\, (\ell^-_v + e_v) \right\} }
\end{eqnarray*}
%
where we find that the collateral price is adjusted by gap risk terms. Examples of gap risk calculation for a CDS can be found in \cite{BrigoCapponiPallavicini}, while for the case of a CCS see \cite{Pallavicini2011}, and references therein.

\subsection{Dealing with Own Credit Risk: Funding Costs and DVA}

Here, we apply our CFBVA master formula to funded instruments, such as an uncollateralized equity option or a corporate bond, which cannot be funded with collateral. We interpret CFBVA prices as the prices calculated by the Investor. We assume that the investor can fund himself at the $f^\pm_t$ rate according to her liquidity policies.

In the present example we recall that expectations are taken under the $\mathbb{Q}^{\tilde f}$ pricing measure, and we assume that no collateral is posted or withdrawn, namely
\[
C_t \doteq 0
\]%
We consider the case where funding and investing operations are entered into on a continuous time-grid. Furthermore, we assume that the close-out amount is calculated as given by
\[
\varepsilon_\tau \doteq \varepsilon_{I,\tau} \doteq \varepsilon_{C,\tau} \doteq \int_\tau^T \ExT{\tau}{\tilde f}{ \Pi(u,u+du) D(\tau,u;{\tilde f}) }
\]%
namely we consider a risk-free close-out amount comprehensive of funding costs. We assume the same funding costs for both the counterparties.

Then, from Theorem \ref{th:fund_hedge} we get
\begin{eqnarray*}
{\bar V}_t(0,F)
&=& \int_t^T \ExT{t}{\tilde f}{ \Pi(u,u+du) \, \exp\left\{ -\int_t^u dv\, {\tilde f}_v \right\}  } \\
& - & \ExT{t}{\tilde f}{ \ind{\tau=\tau_C<T} \lgd_C \varepsilon_\tau^+ \, \exp\left\{ -\int_t^\tau dv\, f^+_v \right\} } \\
& - & \ExT{t}{\tilde f}{ \ind{\tau=\tau_I<T} \lgd_I \varepsilon_\tau^- \, \exp\left\{ -\int_t^\tau dv\, f^-_v \right\} }
\end{eqnarray*}%

According to the examples of liquidity policies we have presented in section \ref{sec:liquidity}, we have two possible choices for investor's funding rate $f^+_t$ and investing rate $f^-_t$.
\begin{enumerate}
\item If the investor is funding by means of his treasury, which applies some form of funds transfer pricing (FTP), we have that the funding/investing rates $f^\pm_t$ are selected by the treasury, and they may depend (indirectly) on the Investor's credit risk.
\item On the other hand, if the investor is able to directly fund himself on the market, then the funding rates $f^\pm_t$ are given by the market itself.
\end{enumerate}

\begin{remark}\label{rem:dvafva}{\bf (FVA and DVA: Deriving earlier results such as \cite{MoriniPrampolini2011} and \cite{Castagna2011} as special cases)}
We consider the simpler case of a positive payoff without first-to-default effect. Thus, only the counterparty's default event is relevant, so that we have only the CVA term. Furthermore, we assume zero recovery rate. We obtain, switching to the default-free market filtration $\cal F$:
\begin{eqnarray*}
{\bar V}_t(0,F)
&=& \ind{\tau>t} \int_t^T \ExT{t}{\tilde f}{ \Pi(u,u+du) D(t,u;f^+) \big| {\cal F} } \\
&-& \ind{\tau>t} \int_t^T du\, \,\ExT{t}{\tilde f}{\lambda_u^C D(t,u;{\lambda^C + f^+}) \int_u^T \ExT{u}{\tilde f}{ \Pi(v,v+dv) D(u,v;f^+) \big| {\cal F} } \big| {\cal F} }\\
&=& \ind{\tau>t} \int_t^T \,\ExT{t}{\tilde f}{ \Pi(u,u+du) \exp\left\{ - \int_t^u dv\, (\lambda^C + f^+) \right\} \big| {\cal F} }
\end{eqnarray*}%
where $\lambda^C_t$ is the counterparty's default intensity.

Then, if we choose the second case above  for the investor's funding rate, so that the investor can adjust the funding rate to include the DVA term when entering the deal opened with the funder (funding benefits), as we described in section \ref{subsec:direct}, we have
\[
f^+_t \doteq e_t + \ell^+_t \,,
\]%
and we obtain a result similar to \cite{MoriniPrampolini2011}. Notice that adjusting the funding rate for funding benefits means that the investor is able to hedge his own credit risk in funding positions. If at this stage one further assumes the funding liquidity spread (basis) $\ell^+$ to be zero, then one has
\[
f^+_t = e_t .
\]%
If one further (and mistakenly) identifies OIS with $r_t$, one has
\[
f^+_t = r_t 
\]
so that one can fund at the risk free rate and there are no funding costs. This is the setup for the claim by \cite{HW2012a}, reported in part of the press, as we pointed out in the introduction, that there is no funding valuation adjustment. We hope it is clear by now under how many special and unrealistic assumptions this happens. 

Going back to $f^+_t \doteq e_t + \ell^+_t$, on the other hand, since we have only the CVA term, we may assume we are dealing only with CVA, so that we can choose not to adjust the funding rate, and we get
\[
f^+_t \doteq e_t + \ell^+_t + \lambda^I_t \,,
\]%
a result usually found in the literature when funding benefits are disregarded or deliberately left aside, see for instance \cite{Castagna2011}.
\end{remark}

\section{Conclusions and Further Developments}

We have seen above that when we try and include consistently funding, credit, (debit) and collateral risk we obtain a highly non-linear and recursive pricing equation. While we derived our equations in the most possible general context allowed by arbitrage freedom, and while we detailed the analysis of funding policies, our findings are in line with other recent findings in the literature, such as \cite{Crepey2011}, \cite{Crepey2012a} and \cite{Crepey2012b}. 

The first outcome from our analysis is that it is difficult to think of funding costs as an adjustment to be added to existing pricing equations, in the sense that something like
{\small\[
AdjustedPrice \quad = \quad RiskFreePrice + DVA - CVA + FVA
\]}%
cannot be obtained in a na\"ive way.
This is highlighted for example in Remark \ref{rem:fpfmdec} for the continuous time case with delta-hedging, but holds more generally for our most general funding-inclusive equations. In the same remark we highlight under which conditions a really  additive decomposition is possible. 
 
%

Our message is that, similarly to how the CVA and DVA analysis made in \cite{BrigoCapponiPallaviciniPapatheodorou} shows that CVA and DVA cannot be considered simply as a spread on discounting curves, in the same way the present paper warns against considering FVA as a mere additive term of the pricing equation, let alone a discounting spread.

Yet, this long journey has not reached its destination, since integrating funding costs into the pricing equations leads to a side effect that has relevant consequences for  the very notion of "price". Indeed, the funding-inclusive price is different for each institution, since each institution has different funding and investing rates depending on her own funding liquidity policy. Even inside the same bank, the treasury and the trading desk may be applying the equation with different inputs. This is why CFBVA prices cannot be agreed upon between two entities. Even collateralized deals, which have an accrual rate defined by the CSA contract, do not have a unique price, since the underlying risk factors grow at a funding rate that can be different for different calculating parties.  Hence a deal price will be reached through negotiation and perhaps an equilibrium approach should be adopted to frame part of the funding costs problem in order to compute the price at which the deal will be actually closed among parties, but this would go beyond the scope of this paper. 

We conclude this paper by stressing the need to go beyond the results we presented here. We have analyzed the mechanics of funding, collateral and hedging to derive the modifications of the general arbitrage-free pricing framework in a way that is consistent with credit and possible debit adjustments. However, we have not yet started analyzing the implications on specific dynamics in financial modelling across asset classes. Most current financial models have to be re-thought to incorporate such notions from scratch. We leave also this task to further work.

A final remark that may be worth considering concerns Central Counterparty Clearing Houses (CCP's). CCPs will not be the end of the credit and funding pricing problem, but just a special case of the general theory we started developing here. A bank may be interested in pricing the same deal under difference scenarios, when clearing it through different CCP's, and seeing how the specific margins she is charged would cover gap risk and wrong way risk. For sensitive deals this can be quite a relevant issue.

\section*{Acknowledgments}

We are grateful to Claudio Albanese, Marco Bianchetti, Enrico Biffis, Cristin Buescu, Antonio Castagna, Lorenzo Cornalba, St\'ephane C\'repey, Cyril Durand, Qing Liu, Nicola Moreni and Giulio Sartorelli for helpful discussions.

\newpage

\bibliographystyle{plainnat}
\bibliography{funding}

\end{document}